\documentclass[12pt]{article}

\usepackage{graphicx}
\usepackage{hyperref}
\usepackage{amsmath,epsfig,epsf,psfrag,natbib}
\usepackage{amssymb}
\usepackage[latin1]{inputenc}
\usepackage{rotating} 
\usepackage{amsbsy}
\usepackage{lscape,url,epstopdf}
\usepackage{color,latexsym,amsfonts,amsthm,amscd,graphicx}
\usepackage{mathrsfs}
 \usepackage{subcaption}
\usepackage{float, multirow, bm, booktabs, threeparttable, diagbox} 
\usepackage[ruled,vlined,linesnumbered]{algorithm2e}

\newcommand{\ben}{\begin{enumerate}}
	\newcommand{\een}{\end{enumerate}} 
\newcommand{\bas}{\begin{eqnarray*}}
	\newcommand{\eas}{\end{eqnarray*}}
\newcommand{\ba}{\begin{eqnarray}}
	\newcommand{\ea}{\end{eqnarray}}
\newcommand{\bit}{\begin{itemize}}
	\newcommand{\eit}{\end{itemize}}

\newtheorem{theorem}{Theorem}

\newtheorem{condition}{Condition}

\usepackage{cleveref}
\usepackage{verbatim}
\usepackage{hyperref}
\usepackage{authblk}
\usepackage{makecell, arydshln}
\usepackage{authblk}

\def\T{{ \mathrm{\scriptscriptstyle \top} }}

\newcommand{\e}{ { \mathbb{E}}}

\newcommand{\argmax}{\mathrm{arg}\mkern1.5mu\mathop{\mathrm{max}}\limits}

\newcommand{\mD}{{\mathcal D}}

\newcommand{\mZ}{{\mathcal Z}}

\newcommand{\bbeta}{ {\bm \beta}}
\newcommand{\btheta}{ {\bm \theta}}

\newcommand{\bu}{ {\bf u}}

\newcommand{\bX}{ {\bf X}}

\newcommand{\bx}{ {\bf x}}

\newcommand{\bTheta}{{\bm \Theta}}

\begin{document}

	\date{}
	\title{
	Revamping Conformal Selection With Optimal Power:   A Neyman--Pearson Perspective
	}
\author[1]{Jing Qin}
\author[2]{Yukun Liu\thanks{Corresponding author:  ykliu@sfs.ecnu.edu.cn}}
\author[3]{Moming Li}
\author[3]{Chiung-Yu Huang}
{\small   \affil[1]{ \small
		National Institute of Allergy and Infectious Diseases, National Institutes of Health}
  \affil[2]{ \small
		KLATASDS-MOE,  School of Statistics,
		East China Normal University,
		Shanghai 200062, China}
  \affil[3]{ \small
		Department of Epidemiology and Biostatistics, University of California at San Francisco, San Francisco, CA 94158, USA}
}
	\maketitle

\abstract{

This paper introduces a novel conformal selection procedure, inspired by the Neyman--Pearson paradigm, to maximize the power of selecting qualified units while maintaining false discovery rate (FDR) control. Existing conformal selection methods may yield suboptimal power due to their reliance on conformal p-values, which are derived by substituting unobserved future outcomes with thresholds set by the null hypothesis. This substitution invalidates the exchangeability between imputed nonconformity scores for test data and those derived from calibration data, resulting in reduced power. In contrast, our approach circumvents the need for conformal p-values by constructing a likelihood-ratio-based decision rule that directly utilizes observed covariates from both calibration and test samples. The asymptotic optimality and FDR control of the proposed method are established under a correctly specified model, and modified selection procedures are introduced to improve power under model misspecification. The proposed methods are computationally efficient and can be readily extended to handle covariate shifts, making them well-suited for real-world applications. Simulation results show that these methods consistently achieve comparable or higher power than existing conformal p-value-based selection rules, particularly when the underlying distribution deviates from location-shift models, while effectively maintaining FDR control.
}


\section{Introduction}
\label{sec:intro}

Conformal prediction has significantly advanced decision-making processes guided by machine learning models. It generates robust prediction sets to quantify prediction uncertainty -- a critical aspect often overlooked in traditional machine learning approaches. Initially formalized by \cite{vovk2005algorithmic}, conformal prediction offers finite-sample coverage guarantees for predictions without relying on distributional assumptions. Subsequent research, including \cite{lei2013distribution}, \cite{lei2018distribution}, \cite{lei2019fast}, \cite{tibshirani2019conformal}, \cite{Romano_2019},  \cite{lei2021conformal}, \cite{chernozhukov_2021}, \cite{barber2021limits},  \cite{Barber_2021_AOS}, and \cite{hu2024two}, has further expanded its capabilities and applicability to various problems.

Beyond its use in prediction, conformal methods have also been applied to selection problems, where the goal is to identify units that meet specific criteria while maintaining control over error rates. This type of conformal method, known as conformal selection, has demonstrated its value in various decision-making and screening contexts. For example, in drug discovery, researchers can leverage conformal inference to prioritize compounds whose prediction intervals indicate a high likelihood of success. This targeted approach allows for a more efficient allocation of resources towards the most promising compounds, thereby optimizing the drug development process. Similarly, when screening job candidates, conformal methods can evaluate the likelihood of success based on qualifications and experience. By focusing on candidates with a high probability of meeting the desired criteria, employers can more effectively allocate resources and streamline the hiring process.

While conformal prediction offers strong marginal coverage guarantees, \cite{JinCandes_2023} and others have highlighted its tendency to produce overconfident selection results. To address this, \cite{JinCandes_2023, JinCandes_2023wt} proposed converting the selection problem as a hypothesis testing task and derive conformal p-values  \citep{bates2023testing} based on the idea of conformal prediction; see, for example, \cite{bao2024selective},
\cite{Gazin2024Selecting},
\cite{marandon2024conformal}
and the references therein for more methodology development and application in various problems. The conformal p-value is the smallest significance level at which a one-sided conformal prediction interval excludes the null hypothesis. To control the false discovery rate, the authors introduced multiplicity control procedures, commonly used in multiple testing scenarios, to ensure that the proportion of falsely selected units among the selected is maintained below a prespecified level. This approach leverages the strengths of conformal inference to provide robust, assumption-free selection decisions.

To our knowledge, all existing conformal selection methods rely on the derivation of conformal p-values as a central element. However, as we will elaborate in the following section, these methods evaluate p-values by substituting unobserved future outcomes in the nonconformity score with thresholds specified by the null hypothesis. This substitution often results in conservative selection results, which can lead to missed opportunities in scenarios where accurately identifying true positives is crucial. These limitations motivate the development of more powerful methods for conformal selection.

To improve the performance of conformal selection, we propose a novel framework that bypasses the computation of conformal p-values. Instead of substituting unobserved outcomes, our approach reformulates the selection problem as a hypothesis test concerning the distribution of observed covariates.
By leveraging the idea of Neyman--Pearson lemma, we construct a likelihood-ratio-based decision rule that directly tests whether the covariate distribution under the null (associated with outcomes not exceeding a threshold) differs from that under the alternative. This reformulation enables the design of selection procedures that achieve asymptotically optimal power while still maintaining strict control over the false discovery rate (FDR). Our contributions are threefold:

\begin{enumerate}
\item \textbf{Direct Testing via Covariate Distribution:} We shift the focus from testing unobserved outcomes to testing the observable covariates. This not only avoids the inherent conservatism of conformal p-value approaches but also permits the construction of valid, direct tests under the Neyman--Pearson paradigm.

\item \textbf{Asymptotically Optimal Power:} By using a likelihood ratio as the selection criterion, our method is shown to be asymptotically optimal in terms of power under the ideal scenario of a correctly specified model. Moreover, we introduce modifications to improve performance under model misspecification.

\item \textbf{Robustness and Flexibility:} Our framework readily extends to settings with covariate shift, accommodating real-world scenarios where the distribution of covariates differs between calibration and test samples. In addition, alternative implementations based on quantile regression further enhance robustness against complex, non-linear relationships.
\end{enumerate}

In summary, while traditional conformal selection methods offer strong finite-sample guarantees, they often sacrifice power due to the conservative nature of conformal p-values. Our work addresses this trade-off by introducing a Neyman--Pearson inspired selection strategy that fully exploits available covariate information, thereby achieving higher power without compromising FDR control. We detail the theoretical properties of our method and demonstrate its superior performance through extensive simulation studies and applications to real-world data.

The rest of this paper is organized as follows. In Section 2, we review existing selection methods based on conformal p-values. In Section 3, we propose a novel conformal selection method based on likelihood ratio in the spirit of the Neyman--Pearson paradigm. In Section 4, we extend our method to handle scenarios under covariate shift and discuss an interesting application to select individuals that can potentially benefit from treatment effect.
Section 5 provides numerical studies of the finite-sample performance of our approaches.
Section 6  offers concluding remarks.
For clarity, all technical proofs are postponed to the supplementary material.

\section{Selection by Prediction With Conformal P-Values}

In this section, we briefly review existing methods for conformal selection. Unlike conformal inference, which focuses on constructing prediction sets with a prespecified coverage guarantee for future outcomes, conformal selection focuses on identifying units that exceed specific thresholds while controlling the error rate. Originally developed based on the principles of conformal prediction, conformal selection has evolved into a powerful tool for facilitating distribution-free decision-making processes.
Suppose we have calibration data $\mD_\textrm{ca}=\{(\bX_1,Y_1),\ldots,(\bX_n,Y_n)\}$ and a test sample $\mD_\textrm{te}=\{(\bX_{n+1},Y_{n+1}),\ldots,(\bX_N,Y_N)\}$, where the test outcomes $Y_{n+1},\ldots,Y_N$ are yet to be observed.
We assume that $\{(\bX_i,Y_i): i=1, \ldots, N\}$ are independently and identically distributed (i.i.d.) replicates of a random vector $(\bX, Y)\in  \mathcal{X} \times \mathcal{Y}$ of an unknown distribution, though later we relax it to accommodate potential distribution shifts in the test sample. For ease of notation, we do not differentiate between random variables and their realizations here. Consider  testing $m=N-n$  one-sided random hypotheses about the future outcomes
$$
H_j:Y_{n+j}\leq c_j \quad
\mbox{versus} \quad
H_j^A:Y_{n+j}> c_j, \quad
j=1,\ldots, m,
$$
where $c_j$ are known thresholds. Here, larger outcome values are of more interest, and each hypothesis corresponds to a specific threshold $c_j$. Note that $c_j$ may vary across units and can be either prespecified or determined randomly.  Unlike conventional hypothesis testing, which focuses on testing population parameters, this setup involves testing individual random outcomes. The goal is to identify as many units satisfying $Y_{n+j}>c_j$ as possible (i.e., rejecting the null hypothesis) while simultaneously controlling the false discovery rate, defined as the expected proportion of false positives among all selected units.

Suppose we have a pretrained prediction model whose training process is independent of both $\mD_\textrm{ca}$ and $\mD_\textrm{te}$. Let $V(\bx, y): \mathcal{X} \times \mathcal{Y} \to \mathbb{R}$ be a nonconformity score function that measures how well an observation $(\bx, y)$ conforms to this model, where $V(\bx,y)$ is monotone in $y$ for any $\bx \in \mathcal{X}$. Popular choices include absolute regression residuals \citep{vovk2005algorithmic}, conformalized quantile regression score \citep{Romano_2019}, and conditional density or cumulative distribution function \citep{chernozhukov_2021} obtained by parametric, nonparametric or machine learning methods. Define $V_i=V(\bX_i,Y_i)$, $i=1,\ldots,n$, for the calibration data and $V_{n+j}=V(\bX_{n+j},Y_{n+j})$, $j=1,\ldots,m$, for the test samples. If $Y_{n+j}$  were available, then $n^{-1}\sum_{i=1}^n I(V_i\le V_{n+j})$ takes values on $\{0, 1/n, 2/n \ldots, 1\}$ with equal probabilities, thus providing a p-value-like measure under the exchangeability condition on the training and test samples. Since $V_{n+j}$ is not evaluable without observing  $Y_{n+j}$, we define $\widetilde V_{n+j}=V(\bX_{n+j},c_j)$.
Then under the null hypothesis $H_j$, we have $V(\bX_{n+j}, Y_{n+j})\le V(\bX_{n+j},c_j) =\widetilde V_{n+j}$ since $V(\bx, y)$ is monotone in $y$. Following \cite{Vovk2015} and \cite{JinCandes_2023}, one can derive a conformal p-value:
\[
\widetilde p_j=\frac{\sum_{i=1}^n I(V_i < \widetilde V_{n+j})+
\{1+\sum_{i=1}^n I(V_i=\widetilde V_{n+j})\} U_j}{n+1},
\]
where $U_j \sim U[0,1]$ are i.i.d. random variables for tie breaking. So $\widetilde p_j$ measures how extreme the threshold $c_j$ is relative to the distribution of $V_i$'s in $\mD_\textrm{ca}$. In fact, as noted in \cite{JinCandes_2023}, $\widetilde p_j$ is the smallest significance level at which  a one-sided conformal prediction interval for $Y_{n+j}$ excludes $c_j$. Hence the use of conformal p-values is equivalent to selection by prediction, much like how conventional p-values relate to confidence intervals.

As anticipated, applying conformal p-value to test multiple random hypotheses leads to an inflated error rate.
To address this, \cite{JinCandes_2023} proposed applying the Benjamini--Hochberg procedure \citep{BH_1995} to control the false discovery rate, that is, the expected proportion of false positives among all positive selections. However, it is important to note that, under the null hypothesis and given the monotonicity of $V$, we have $V_{n+j}\le \widetilde V_{n+j}$ and hence $I(V_i\le V_{n+j})\le I(V_i\le \widetilde V_{n+j})$. This implies  $\widetilde p_j\ge p_j $, where $p_j $ is the oracle conformal p-value if $Y_{n+j}$ were available. Specifically,
\[
p_j =\frac{\sum_{i=1}^n I(V_i < V_{n+j})+
\{1+\sum_{i=1}^n I(V_i=V_{n+j})\} U_j}{n+1},
\]
and thus $p_j $ measures how extreme $V_{n+j}$, rather than $\widetilde  V_{n+j}$, is relative to the distribution of $V_i$'s in $\mD_\textrm{ca}$. Thus the conformal p-values tend to yield more conservative results than the oracle ones. Consequently, applying the Benjamini--Hochberg procedure to the conformal p-values to control FDR is likely to result in low power in selecting units that meet the criteria, thus compromising the efficiency of the selection process.

Determining the optimal nonconformity score function is generally challenging. For binary outcomes, \cite{JinCandes_2023} proposed using a clipped score $V(\bx, y)=MI(y>c) + c I(y\leq c) - \widehat\mu(\bx)$ to achieve better power while controlling for FDR. Here, $M$ is a sufficiently large constant, and $\widehat{\mu}(\bx)$ is an estimator of  $\Pr(Y=1\mid\bX=\bx)$. However, their method does not readily extend to continuous outcomes. In this paper, we demonstrate that it is possible to directly control the FDR and maximize the power function without computing conformal p-values, even for outcomes with arbitrary distributions.

\section{Conformal Selection With Optimized Power}

\subsection{The Proposed Selection Procedure}
\label{constant_c}

We propose a novel conformal selection procedure that bypasses the derivation of conformal p-values and can achieve optimal power with any given nonconformity score. Existing conformal selection methods address the issue of unobserved test outcomes $Y_{n+j}$ by substituting them with the prespecified thresholds $c_j$ to obtain  $\widetilde  V_{n+j}$. The null hypothesis is then rejected when $\widetilde  V_{n+j}$ is too extreme relative to the nonconformity scores $V_i$ obtained from the calibration data. This strategy is somewhat indirect because it relies on substituting unobserved outcomes with the thresholds specified by the null hypothesis. In contrast, we propose a more direct method by testing the distribution of the observed covariates under the null hypothesis. By focusing on the covariate distribution, we can fully leverage the available information without the need to impute or replace unobserved outcomes.

We begin by assuming that the threshold is constant across testing units, that is,  $c_j=c$, $j=1,\ldots, m$, and that the calibration and test samples were drawn from the same distribution. Let $g(\bx)$ denote the probability density function of the covariate vector $\bX$. Denote the conditional cumulative distribution function of $Y$ given $\bX$ as $F(y\mid \bx)$, and define the corresponding conditional survival function as $\bar  F(y\mid \bx)=1-F(y\mid \bx)$. Similarly, let $F_Y(y)$ denote the marginal cumulative distribution function of $Y$, and define the marginal survival function as $\bar  F_Y(y)=1-F_Y(y)$.  Under the null hypothesis that $H_j:Y_{n+j}\leq c_j$, the conditional density function of $\bX_{n+j}$ given $Y_{n+j}\leq c$ is
\[
g_0(\bx; c) := \frac{F(c \mid \bx)g(\bx)}{F_Y(c)},
\]
while analogously the conditional density function of $\bX_{n+j}$ given $Y_{n+j}> c$ under the alternative hypothesis is
\[
 g_1(\bx; c) := \frac{\bar {F}(c \mid \bx)g(\bx)}{\bar {F}_Y(c)}.
\]
In this way, we convert the problem of testing the random hypothesis  $Y_{n+j}\leq c$ against  $Y_{n+j}> c$  into testing the conventional-type null hypothesis $\bX_{n+j} \sim g_0(\bx; c)$ against $\bX_{n+j} \sim g_1(\bx; c)$. In other words, this reformulation shifts from testing an unobserved random outcome to testing the distribution of observed covariates. This approach allows us to leverage information from observed covariates to test hypotheses about the unobserved outcomes, facilitating the application of conventional hypothesis testing methods.
Note that since $Y_{n+j}$ in the test sample is not observed, $\bX_{n+j}$ can be viewed as a mixture of two random variables with density functions $g_0(\bx; c)$ and $g_1(\bx; c)$ and mixing proportions $F_Y(c)$ and $\bar  F_Y(c)$, respectively; more specifically, the marginal density function of $\bX_{n+j}$ can be expressed as a mixture of two density functions, that is, $g(\bx)=F_Y(c)g_0(\bx; c) + \bar  F_Y(c) g_1(\bx; c)$.

Inspired by the Neyman--Pearson paradigm, we consider the likelihood ratio
\[
R^*(\bx; c):=\frac{g_0(\bx; c)}{g_1(\bx; c)}=\frac{F(c \mid \bx)}{\bar {F}(c \mid \bx)}\frac{\bar {F}_Y(c)}{F_Y(c)}.
\]
Then rejecting the one-sided null hypothesis $H_j$ when $R^*(\bX_{n+j}; c)$ is small is expected to yield the most powerful test. Note that the rejection region $ \{ R^*(\bx; c) \leq \eta^*\}$ is equivalent to  $ \{  R(\bX_{n+j}; c) \leq \eta \}$ with $\eta=\eta^* F_Y(c) / \bar {F}_Y(c)$ and
\[
R(\bx;c): =\frac{F(c \mid \bx)}{\bar {F}(c \mid \bx)}.
\]
Denote by $\mathcal{S}=\{j: R(\bX_{n+j}; c) \leq \eta, j=1,\ldots, m \}$ the collection of selected units. To control the error rate in decision-making, we aim to set $\eta$ to control FDR, which is the expected false discovery proportion (FDP) of the selected units:
\begin{equation*}
     \mbox{FDR} (\eta;c) = \e\left[\frac{\sum_{j=1}^m I\{ R(\bX_{n+j}; c) \leq \eta, Y_{n+j}\le c\} }{1\vee \sum_{j=1}^m  I\{ R(\bX_{n+j}; c) \leq \eta\} }\right],
\end{equation*}
where $s \vee t = \max\{s,t\}$.
At the same time, we wish to select a critical value $\eta$ that maximizes the power, that is, the expected proportion of positive units being selected
\begin{equation*}
 \mbox{Power} (\eta;c) = \e\left[\frac{\sum_{j=1}^m I\{ R(\bX_{n+j}; c) \leq \eta, Y_{n+j}> c\} }{1\vee \sum_{j=1}^m  I(Y_{n+j}> c)  } \right].
\end{equation*}

Note that both FDR$(\eta;c) $ and Power$(\eta;c) $ can not be evaluated directly with the test sample as $Y_{n+j}$ is unavailable. When $m$ tends to infinity, we have
$$
\lim_{m\rightarrow\infty} \mbox{FDR} (\eta;c) = \frac{ \Pr\{R(\bX; c) \leq \eta, Y\le c\}}{\Pr\{R(\bX; c) \leq \eta\}}  \;  \mbox{and}\;
\lim_{m\rightarrow\infty} \mbox{Power} (\eta;c) = \frac{ \Pr\{R(\bX; c) \leq \eta, Y> c\}}{\Pr(Y>c)}.
$$
Obviously, with the calibration dataset, the two quantities can be estimated by their empirical version
\allowdisplaybreaks\begin{align}
\label{zeta-n}
\zeta_{n}(\eta;c)&= \frac{ \sum_{i=1}^n I\{R(\bX_i;c) \leq \eta, Y_i \leq c \}}{ 1 \vee \sum_{i=1}^n I\{R(\bX_i;c) \leq \eta\}} \quad  \mbox{and} \\
\label{Psi-n}
\Psi_{n}(\eta;c) &= \frac{ \sum_{i=1}^n I\{R(\bX_i;c) \leq \eta, Y_i >c \}}{  1 \vee \sum_{i=1}^n I(Y_i> c) },
\end{align}
respectively.
Then $\eta$ is determined by maximizing $\Psi_n(\eta;c)$ subject to $\zeta_n(\eta;c)\le a$, where $a$ is the FDR target level, and we reject $H_j$ when $R(\bX_{n+j}; c)\le \eta$, $j=1, \ldots, m$. It is worthwhile to point out that identifying the optimal $\eta$ is computationally straightforward. Specifically, both $\zeta_n(\eta;c)$ and $\Psi_n(\eta;c)$ are step functions with jumps at $\eta_i=R(\bX_i; c)$, $i=1, \ldots, n$. Therefore, evaluating these functions requires only computing their values at these discrete points. Let $E_n =\{\eta_i: \zeta_n(\eta_i; c)\le a, 1\leq i \leq n\}$ represent a candidate set of critical values that satisfy the FDR constraint.  The optimal critical value $\eta_n^{\rm opt}$ is then set as the value in $E_n$ that maximizes the power function, i.e.,
 $$
\eta_n^{\rm opt}=\argmax_{\eta\in E_n} \Psi_n(\eta; c).
$$
Since the power function is increasing in $\eta$, the optimal $\eta$ can be further simplified to $\eta_n^{\rm opt}=\max\{\eta_i: \zeta_n(\eta_i; c)\le a, 1\leq i \leq n\}$. It is easy to see that the proposed selection method is flexible enough to accommodate alternative definitions of the power function.

Interestingly, in the special case of location-shift models, the proposed selection method performs comparably to the conformal p-value approach with a clipped score \citep{JinCandes_2023}. To see this, consider a pretrained model $Y=\mu(\bX)+\epsilon$, where the random error $\epsilon$ has a CDF $F_\epsilon$. It can be verified that $R(\bX;c) \leq \eta$ is equivalent to $\mu(\bX) \ge c- F_\epsilon^{-1}(\eta/(1+\eta))$. So the proposed selection algorithm is equivalent to ranking $\mu(\bX_{n+j})$ alongside $\{\mu(\bX_{i})\}_{i=1}^n$. On the other hand, the clipped score is given by $MI(y>c) + c I(y\leq c) -\mu(\bx)$, where $M$ is a sufficiently large number satisfying  $M\ge 2\sup_{\bx} |\mu(\bx)|$. This choice of $M$ ensures that the nonconformity score for units with $\{Y>c\}$ is always greater than that for units with $\{Y\le c\}$. Furthermore, the ranking of units with $\{Y\le c\}$ depends solely on their $\mu(\bX_i)$ values. Following the argument in Section 2.5 of \cite{JinCandes_2023}, this setup ensures that the FDR approaches the desired nominal level. Given that our method also ranks $\mu(\bX_i)$'s, it naturally achieves the same result, ensuring that the FDR approaches the desired nominal level while maximizing power. In the next subsection, we formally establish the optimality of our proposed selection method.

\subsection{Asymptotic Optimality}

The preceding discussion assumes that $F(y\mid\bx)$ is a pretrained model. In practice, a parametric distribution, say $F(y\mid\bx;\btheta)$, can be fit as a working model for $Y$ given $\bX$, with the unknown parameter $\btheta$ estimated using a training dataset. A common choice for continuous outcomes is the location-shift model $Y=\mu(\bX; \btheta)+\epsilon$, where $\epsilon$ follows a mean-zero normal distribution. Here the mean function $\mu(\cdot)$ can be either prespecified or estimated using machine learning techniques like random forests and neural networks, though the latter typically incur higher computational costs due to their complexity.

Let $R_n(\bX;\btheta,c)$, $\zeta_n(\eta;\btheta,c)$, and $\Psi_n(\eta;\btheta,c)$ be the counterparts of $R_n(\bX;c)$, $\zeta_n(\eta;c)$, and $\Psi_n(\eta;c)$, respectively, with $F(c\mid\bX)$ replaced with $F(c\mid\bX; \btheta)$. Algorithm \ref{algo_c_no_wt} outlines the proposed selection procedure. In Theorem \ref{thm-fdr-nc-cc}, we show that, under Condition \ref{condition-F}, the proposed method asymptotically controls the FDR.

\begin{center}
\begin{algorithm}[h!]
\caption{Conformal Selection with a Constant Threshold in the Absence of Covariate Shift}
\label{algo_c_no_wt}

\KwIn{Calibration data $\mD_\textrm{ca} = \{ (\bX_i, Y_i): 1\leq i \leq n \}$, test covariate data $\{\bX_{n+j}: 1\leq j \leq m\}$, a threshold $c$, target FDR $ a \in (0,1)$, a parametric working model $F(y\mid \bx; \btheta)$, an error function $\zeta_n(\eta;\btheta, c)$, a power function $\Psi_n(\eta;\btheta, c)$, and an estimate $\widehat{\btheta}_n$ for $\btheta$ pretrained on a training set}

\KwOut{Selection set $\mathcal{S}$}
For $1\leq i\leq n$, compute $\eta_i = R(\bX_i;\widehat{\btheta}_n, c) := F(c\mid \bX_i;\widehat{\btheta}_n)/\{1-F(c\mid \bX_i;\widehat{\btheta}_n)\} $.

Obtain a set of candidate critical values $E_n = \{\eta_i: \zeta_n(\eta_i;\widehat{\btheta}_n, c) \leq  a, 1\leq i \leq n\}$.

If $E_n \neq \varnothing$, set $\eta_n^{\rm opt} = \argmax_{\eta \in E_n} \Psi_n(\eta; \widehat{\btheta}_n, c)$; otherwise, set $\eta_n^{\rm opt}= -1 $.

\KwResult{A conformal selection set is $\mathcal{S}=\{j: R(\bX_{n+j};\widehat{\btheta}_n, c) \leq \eta_n^{\rm opt},  1\leq j\leq m \}$ }
\end{algorithm}
\end{center}
\vspace{-1.5cm}

\begin{condition}
\label{condition-F}
(i)
$F(y\mid \bx; \btheta)$ is continuous with respect to
$\btheta$ in a compact set $\bTheta$;
(ii)
$\widehat{\btheta}_n \in \bTheta$ converges in probability to  an element $\btheta^*$ in  $\bTheta$; and
(iii)
 $ \sup\limits_{(\eta, \btheta) \in [0,1]\times \bTheta}\Pr\{ F(c\mid \bX; \btheta)  = \eta \} = 0$ for any given $c$.
\end{condition}

Condition \ref{condition-F} allows $\btheta$ to take values in both Euclidean and non-Euclidean spaces, including vector-valued spaces and function spaces. Thus it accommodates estimates derived using nonparametric and machine learning methods.
Condition \ref{condition-F}(iii) excludes situations where, as a random variable, $F(c \mid \bx; \btheta)$ has a positive probability mass for some $c$ and $\btheta$, thus ensuring that $\zeta_n(\eta;\btheta,c)$ converges to $\lim\limits_{m\rightarrow\infty} \mbox{FDR} (\eta;c)$. Without this condition, convergence may be compromised, potentially undermining asymptotical FDR control.

\begin{theorem}
\label{thm-fdr-nc-cc}
Suppose the calibration data $\mathcal{D}_{\rm ca}$ and the test data $\mathcal{D}_{\rm te}$ are i.i.d., and that Condition \ref{condition-F} holds.  Given $ a \in (0, 1)$,  let  $ \mathcal{S}$ be the output of Algorithm \ref{algo_c_no_wt}.
Then
 $$
\limsup_{n\rightarrow \infty}\e\left[
\frac{ \sum_{j=1}^m I(j \in \mathcal{S}, Y_{n+j} \leq c )}{ 1 \vee |\mathcal{S}|}
\right] \leq a.
 $$
If, in addition, $\eta_n^{\rm opt}$ converges in probability to a constant $\eta^{\rm opt}$, then we have
$$
\lim_{n\rightarrow \infty}   \e\left[
\frac{ \sum_{j=1}^m I(j \in \mathcal{S}, Y_{n+j} \leq c )}{ 1 \vee |\mathcal{S}|}
\right]
=
a \times
 \{1 - (1-b)^{m}  \},
$$
where
$b =\Pr\{   R(  \bX ; \btheta^*, c ) \leq \eta^{\rm opt}   \} $.
\end{theorem}

It is important to point out that Condition \ref{condition-F} does not require $F(y\mid\bx; \btheta)$ to be correctly specified. Therefore, by Theorem \ref{thm-fdr-nc-cc}, the proposed selection procedure maintains asymptotic FDR control even when the working model $F(y\mid\bx; \btheta)$ is misspecified.
Moreover, Theorem \ref{thm-fdr-nc-cc} also implies that the asymptotic FDR achieved is slightly below the target level $a$, with the difference diminishing quickly as $m$ increases. However, as suggested by Theorem \ref{thm-fdr-nc-cc-power} below, our method can fully utilize the FDR level by employing an adjusted target FDR rate.
Interestingly, when $F(y\mid\bx; \btheta)$ is correctly specified and the FDR level is fully utilized, the proposed selection procedure achieves asymptotic optimality in terms of power; see Theorem \ref{thm-fdr-nc-cc-power} below. This result mirrors the optimality of the likelihood ratio test, as established by the well-known Neyman--Pearson lemma.

\begin{theorem}
\label{thm-fdr-nc-cc-power}
Suppose the calibration data $\mathcal{D}_{\rm ca}$ and test data $\mathcal{D}_{\rm te}$ are i.i.d., and that Condition \ref{condition-F} holds. Let  $a \in (0, 1)$ be a prespecified FDR level, and let $ \mathcal{S}$ be the output of Algorithm \ref{algo_c_no_wt} with   FDR level
$
\widetilde a = a
\{1 -   ( 1- \widehat b )^{m} \}^{-1},
$
where
$
\widehat b = n^{-1}\sum_{i=1}^n I\{  R(\bX_i; \widehat{\btheta}_n, c) \leq \eta_n^{\rm opt}   \}.
$
Assume that $F(y\mid\bx; \btheta)$ is correctly specified and that the power function $\Psi_n(\eta;\btheta, c)$ in Algorithm \ref{algo_c_no_wt}  increases with $\eta$ for each $\btheta$ and $c$. Then, as $n\rightarrow \infty$, the proposed selection method is asymptotically the most powerful in maximizing the proportion of units that meet the criteria to be selected, among all rules with FDR asymptotically controlled at or below $a$.
\end{theorem}

Note that Theorem \ref{thm-fdr-nc-cc} only guarantees asymptotic FDR control. To achieve finite-sample FDR control, we propose using a modified error function
\begin{equation}
\label{zeta-n-plus}
\zeta_n(\eta;c,\delta)= \frac{ \delta + \sum_{i=1}^n I\{R(\bX_i;c) \leq \eta, Y_i \leq c \} }{ 1 \vee \sum_{i=1}^n I\{R(\bX_i;c) \leq \eta\}},
\end{equation}
where $\delta\geq 0$ serves as a tuning parameter that adjusts the strictness of FDR control. This strategy has been utilized by other authors, including \cite{barber2015controlling} in their Knockoff+ method, to ensure robust FDR control. Our simulation studies suggest that setting $\delta \in \{0.5,1.0\}$ effectively balances FDP and power, offering a practical solution for finite-sample selection scenarios.

\subsection{Maximizing Power Under Model Misspecification \label{max-power-model-mis}}

In practice, the postulated parametric model may not be correctly specified, potentially leading to suboptimal performance. To address this issue, we propose treating the finite-dimensional $\btheta$ in the working parametric model
$ F(y\mid \bx; \btheta)$ as a free parameter rather than fixing its value according to a pretrained model.
Specifically, write $R(\bx; \btheta, c)= F(c\mid \bx; \btheta)/\bar  F(c\mid \bx; \btheta)$.
We maximize the power function
\begin{equation}
\label{Psi-n-theta-eta}
\Psi_n(\eta, \btheta;c) = \frac{\sum_{i=1}^n I\{R(\bX_i; \btheta, c) \leq \eta, Y_i > c\} }{1\vee \sum_{i=1}^n  I(Y_i > c) }
\end{equation}
with respect to $\eta$ and $\btheta$ simultaneously, subject to the constraint
\begin{equation}
\label{zeta-n-theta-eta}
\zeta_n(\eta, \btheta;c) = \frac{\sum_{i=1}^n  I\{ R(\bX_i; \btheta, c) \leq \eta, Y_i \leq c \}}{1\vee \sum_{i=1}^n I\{R(\bX_i; \btheta, c) \leq \eta\} } \leq a.
\end{equation}
Denote by $(\widehat \btheta_n^{\rm opt}, \widehat \eta_n^{\rm opt})$ the solution to the optimization problem described above.  Alternatively, we may obtain estimates by applying a two-step procedure. First, we calculate $ \eta(\btheta) = \arg\max_{\eta} \Psi_n(\eta, \btheta;c)$ subject to $\zeta_n(\eta, \btheta;c) \leq a$ for a fixed $\btheta$. We then set $\widehat \btheta_n^{\rm opt}= \argmax_{\btheta\in\bTheta}\Psi_n(\eta(\btheta), \btheta;c)$ and $\widehat \eta_n^{\rm opt} = \eta(\widehat \btheta_n^{\rm opt})$. 
Note that when the feature space is of high dimension, a possible strategy is to reduce the dimensionality of $\bX$ using variable selection methods such as LASSO \citep{Tibshirani_1996_LASSO} and SIS \citep{FanLv_2008_SIS}, and then implement  the above selection procedure
with the selected predictors.

While the focus here is not on parameter estimation, it is worth noting that if the model $F(y\mid\bx; \btheta)$ is correctly specified with the true parameter value $\btheta^*$, then $\widehat{\btheta}_n^{\rm opt}$ is consistent for $\btheta^*$, as stated in Theorem \ref{thm-op-nc-cc}. By a similar argument to that in the proof of Theorem  \ref{thm-fdr-nc-cc}, the FDR of the  conformal selection procedure
remains asymptotically controlled.

\begin{theorem}
\label{thm-op-nc-cc}
Suppose the calibration data $\mathcal{D}_{\rm ca}$ and the test data $\mathcal{D}_{\rm te}$ are i.i.d., and that Condition \ref{condition-F}  holds. Assume that $F(y\mid\bx; \btheta)$ is a correctly specified model for $\Pr(Y\leq y\mid\bX=\bx)$ with
a finite-dimensional parameter $\btheta$ and  $\btheta^*$ its true  value. Suppose $\btheta$ is identifiable, that is, $\Pr\{F(c\mid \bX ; \btheta_1) \neq F(c\mid \bX ; \btheta_2 ) \} >0$ for any $\btheta_1 \neq \btheta_2$ and $\btheta_1, \btheta_2 \in \bTheta$. Let  $a \in (0, 1)$ be a prespecified FDR level and  $ (\widehat{\btheta}_n^{\rm opt}, \widehat{\eta}_n^{\rm opt})$  be the maximizer of  \eqref{Psi-n-theta-eta} subject to \eqref{zeta-n-theta-eta}. Then $\widehat{\btheta}_n^{\rm opt} = \btheta^* + o_p(1)$  as $n\rightarrow \infty$.

\end{theorem}

\subsection{Conformal Selection With Varying Thresholds \label{subsection-vc}}

The procedure outlined in Algorithm \ref{algo_c_no_wt} is tailored for the situation where $c_j\equiv c$ for all $j=1, \ldots, m$. However, in many applications, the thresholds $c_j$ may vary across units. In such cases, both the error function and the power function depend on the individual values of $c_j$, and applying the same selection rule across all units can lead to suboptimal performance. To account for these variations, we propose a modified algorithm that applies a unit-specific selection rule to handle the varying thresholds while achieving FDR control. Specifically, for each $j=1, \ldots, m$, we maximize the power function $\Psi_n(\eta; c_j)$ with respect to $\eta$, subject to $\zeta_n(\eta; c_j)\leq  a $ where $\zeta_n$ and $\Psi_n$ are defined in \eqref{zeta-n} and \eqref{Psi-n}, respectively. Write $\eta_{ij} = R(\bX_i; c_j)$, $i = 1, \ldots, n$. As argued in Section \ref{constant_c}, we obtain a set of candidate critical values $E_{nj} = \{\eta_{ij}: \zeta_n(\eta_{ij};  c_j) \leq  a, 1\leq i \leq n\}$ and then identify the optimal unit-specific critical value  $\eta_{nj}^{\rm opt} = \argmax_{\eta \in E_{nj}} \Psi_n(\eta;  c_j)$. The proposed conformal selection set is then given by $\mathcal{S} = \{j:  R(\bX_{n+j};c_j) \leq \eta_{nj}^{\rm opt},\;  1\leq j\leq m \}$. Details can be found in Algorithm \ref{algo_cj_no_wt}.

\begin{center}
\begin{algorithm}[h!]
\caption{Conformal Selection With Varying Thresholds in the Absence of Covariate Shift}
\label{algo_cj_no_wt}

\KwIn{Calibration data $\mD_\textrm{ca} = \{ (\bX_i, Y_i): 1\leq i \leq n \}$, test covariate data $\{\bX_{n+j}: 1\leq j \leq m\}$, thresholds $\{c_j\}_{j=1}^m$, target FDR $a \in (0,1)$, a parametric working model $F(y\mid \bx; \btheta)$, an error function $\zeta_n(\eta;\btheta, c)$, a power function $\Psi_n(\eta;\btheta, c)$, and an estimate $\widehat{\btheta}_n$ for $\btheta$ pretrained on a training set}

\KwOut{Selection set $\mathcal{S}$}

\For{$j=1$ \KwTo $m$} {

For $1\leq i \leq n$, obtain $\eta_{ij} = R(\bX_i;\widehat{\btheta}_n, c_j):= F(c_j\mid \bX_i;\widehat{\btheta}_n)/\{1-F(c_j\mid \bX_i;\widehat{\btheta}_n)\}$.

Obtain a candidate  set $E_{nj} = \{\eta_{ij}: \zeta_n(\eta_{ij};\widehat{\btheta}_n, c_j) \leq  a, 1\leq i \leq n\}$.

If $E_{nj} \neq \varnothing$, find $\eta_{nj}^{\rm opt} = \argmax_{\eta \in E_{nj}} \Psi(\eta;\widehat{\btheta}_n, c_j)$; otherwise set $\eta_{nj}^{\rm opt}  =   -1 $.

}

\KwResult{A conformal selection set is $\mathcal{S}=\{j: R(\bX_{n+j};\widehat{\btheta}_n,c_j) \leq \eta_{nj}^{\rm opt} ,  1\leq j\leq m \}$ }
\end{algorithm}
\end{center}

In Theorem \ref{thm-fdr-nc-vc} below,  we show that our conformal selection procedure with varying thresholds asymptotically control FDR, and, with slight modification, its limit FDR achieves exactly the target level.  Furthermore, after  modification, if the working model $F(y\mid\bX; \btheta)$ is correct,  asymptotically it enjoys an optimality in terms of power; see Theorem  \ref{thm-fdr-nc-vc-power}.

\begin{theorem}
\label{thm-fdr-nc-vc}
Suppose that the calibration data $\mathcal{D}_{\rm ca}$ and the test data $\mathcal{D}_{\rm te}$ are i.i.d., and that Condition  \ref{condition-F} holds.
  In addition, assume that the varying thresholds $c_j$, $j=1, \dots, m$, are nonrandom constants or random variables that are independent of $\bX_{n+j}$.
Then, for $ a \in (0, 1)$,  the output of Algorithm \ref{algo_cj_no_wt} satisfies
 $$
\limsup_{n\rightarrow \infty}\e\left[
\frac{ \sum_{j=1}^m I(j \in \mathcal{S}, Y_{n+j} \leq c_j )}{ 1 \vee |\mathcal{S}|}
\right] \leq a.
$$
Further, if $\eta_{nj}^{\rm opt} $ in  Algorithm \ref{algo_cj_no_wt}  converges to  constants $\eta_{j}^{\rm opt} $ for  $1\leq j\leq m$ as $n\rightarrow \infty$,
then
  $$
\lim_{n\rightarrow \infty}\e\left[
\frac{ \sum_{j=1}^m I(j \in \mathcal{S}, Y_{n+j} \leq c_j )}{ 1 \vee |\mathcal{S}|}
\right]  = a\times \left\{ 1- \prod_{j=1}^m (1-b_j) \right\},
$$
where $b_j =\Pr\{   R(  \bX ; {\btheta^*}, c_j ) \leq \eta_j^{\rm opt}   \} $.
\end{theorem}

Under mild conditions, $E_{nj}$ converges to a closed and bounded set
$E_j$ as $n\rightarrow\infty$ (see the proof of Theorem \ref{thm-fdr-nc-vc}). In general,   for a fixed $c_j$, the error function $\zeta_n(\eta, \widehat \btheta; c_j) $
converges in probability to
$L(\eta) =  \Pr_{\rm tr}\{  Y  \leq c_j\mid R(\bX; \btheta_*, c_j) \leq \eta\} $.
Additionally, suppose that  $L(\eta)$ is continuous. This together with $0\leq L(\eta) \leq 1$ implies that  the maximizer
$\eta_j^{\rm opt} $ of $L(\eta)$ must exist. If the convergence of $\zeta_n(\eta, \widehat \btheta; c_j) $ to $L(\eta)$ can be strengthened to uniform convergence over $\eta\in E_j$ and the maximizer
$\eta_j^{\rm opt} $ of $L(\eta)$ is unique, then we have  $\eta_{nj}^{\rm opt}  = \eta_j^{\rm opt} + o_p(1)$.

Similar to Theorem \ref{thm-fdr-nc-cc}, Theorem \ref{thm-fdr-nc-vc} implies that the asymptotic FDR achieved by the proposed selection procedure is slightly below the target level, with the difference diminishing rapidly as $m$ increases. Moreover, implementing the selection procedure with an adjusted target FDR level  allows for full utilization of the FDR allowance, thereby potentially improving power.
Theorem \ref{thm-fdr-nc-vc-power} below establishes that the proposed selection procedure, when implemented with an adjusted target FDR level, yields the most powerful selection rule among methods that control FDR at the same target level for each test sample.

For a test unit $j$, one can only observe the covariate $\bX_{n+j}$. Thus, the resulting selection set must be of the form $\mathcal{T} = \{j:  \bX_{n+j} \in T_{nj},  1\leq j\leq m \}$. Let $a \in (0, 1)$ be a prespecified FDR level and
\begin{equation}
\label{adjusted-fdr}
\widetilde a_1 = a
\left\{1 -   \prod_{j=1}^m ( 1- \widehat b_j )  \right\}^{-1},
\end{equation}
where
$
\widehat b_j = n^{-1}\sum_{i=1}^n I\{  R(   \bX_i ; \widehat \btheta_n, c_j    ) \leq \eta_{nj}^{\rm opt}   \}.
$
Let $\mathscr{T}$ denote the set of all the selection rules at the FDR level $\widetilde a_1$ whose selection sets can be expressed as
$ \mathcal{T}  = \{j:  \bX_{n+j}  \in T_{nj},  1\leq j\leq m \}$ with $T_{nj}$ satisfying
$
  \sum_{i=1}^n  I\{ \bX_{i}  \in T_{nj} ,    Y_{i} \leq c_j \} /[  1 \vee\sum_{i=1}^n
I\{\bX_{i}  \in T_{nj} \}]   \leq  \widetilde a_1.
$

\begin{theorem}
\label{thm-fdr-nc-vc-power}
Suppose that the calibration data $\mathcal{D}_{\rm ca}$ and the test data $\mathcal{D}_{\rm te}$ are i.i.d., and that Condition  \ref{condition-F} hold.
Let $ \mathcal{S}$ be the output of Algorithm \ref{algo_cj_no_wt} at FDR level
$\widetilde a_1$ in \eqref{adjusted-fdr}.
 Assume that     the varying thresholds  $c_j$, $j=1, \dots, m$, are  nonrandom constants or   random variables that are independent of $\bX_{n+j}$.
In addition, assume that $F(y\mid\bx;\theta)$ is correctly specified for $\Pr(Y\leq y \mid \bX = \bx)$ and the power function  $\Psi(\eta;\btheta, c)$ in Algorithm  \ref{algo_cj_no_wt} is an increasing function of $\eta$.
Then, as $n\rightarrow \infty$, the proposed selection method is asymptotically the most powerful in maximizing the proportion of units that satisfy the prespecified criteria being selected, among all selection rules in $\mathscr{T}$.
\end{theorem}

\subsection{Conformal Selection Based on Quantile Regression}

In previous sections, we showed that the asymptotic FDR control of the proposed selection method is robust against misspecifications of $F(y\mid \bx)$. However, the asymptotic optimality of the likelihood-ratio-based method relies on having a correctly specified CDF. While simple parametric working models for $F(y\mid \bx)$ can be employed in practice, they may fail to to capture complex nonlinear or heteroscedastic relationships in real-world data. A key observation is that the rejection set ${R(\bx;c) \leq \eta}$ can be reexpressed as ${c \leq F^{-1}(\eta^\dag \mid \bx)}$ for some $\eta^\dag$, motivating the use of quantile regression as a working model. Quantile regression \citep{koenker1978regression, koenker2005quantile} extends traditional linear regression by modeling quantiles of the outcome distribution. Moreover, our simulation studies show that quantile regression often outperforms simple regression models in settings with non-normality, heteroscedasticity case, further supporting its use. To simplify the discussion, we use $\eta$ and $\eta^\dag$ interchangeably in the rest of the paper, as this does not affect the analysis.

We propose to construct the rejection set $\{c\le \widehat Q(\eta^{\rm opt}\mid \bX)\}$ with $\eta^{\rm opt}$ being the largest quantile  $\eta \in \{t_l\}_{l=1}^L$ that maximizes the power function
\begin{equation} \label{Psi_Q}
\Psi_n^Q(\eta; c)
= \frac{\sum_{i=1}^n I\{ c \leq \widehat{Q}(\eta\mid\bX_i), Y_i > c\}}{1 \vee \sum_{i=1}^n I(Y_i > c)}
 \end{equation}
while controlling the error function under a prespecified nominal level $a$, that is,
 \begin{equation} \label{zeta_Q}
\zeta_n^Q(\eta; c) = \frac{\sum_{i=1}^n I\{c \leq \widehat{Q}(\eta\mid\bX_i), Y_i \leq c\}}{1 \vee \sum_{i=1}^n I\{c \leq \widehat{Q}(\eta\mid\bX_i) \}}
\leq a.
 \end{equation}
The procedure can be readily extended to handle cases with varying thresholds $c_j$, as described in Section 3. Algorithm \ref{algo_cj_no_wt_qrf} outlines the implementation of our selection method based on quantile regression with varying $c_j$. To establish the asymptotic properties of the proposed conformal selection approach, we impose the following condition.

\begin{center}
\begin{algorithm}[h!]
\caption{ Conformal Selection For Varying Thresholds in the Absence of Covariate Shift Using Quantile Regression Working Model}
\label{algo_cj_no_wt_qrf}

\KwIn{Calibration data $\mD_\textrm{ca} = \{ (\bX_i, Y_i): 1\leq i \leq n \}$, testing covariate data $\{\bX_{n+j}: 1\leq j \leq m\}$, thresholds $\{c_j\}_{j=1}^m$, target FDR $a \in (0,1)$, a pretrained conditional quantile model $\widehat{Q}(t\mid\bX)$, candidate quantiles $\{t_l\}_{l=1}^L$, the error function $\zeta_n^Q(\eta;c)$ in \eqref{zeta_Q}, and the power function $\Psi_n^Q(\eta;c)$ in \eqref{Psi_Q}}

\KwOut{Selection set $\mathcal{S}$}

\For{$j=1$ \KwTo $m$} {

Obtain a candidate set $E_{nj} = \{t_l: \zeta_n^Q(t_l;c_j) \leq a, 1\leq l \leq L\}$.

If $E_{nj} \neq \varnothing$, find $\eta_{nj}^{\rm opt} = \argmax_{\eta \in E_{nj}} \Psi_n^Q(\eta;c_j)$; otherwise set $\eta_{nj}^{\rm opt} = t_1/2$.

}

\KwResult{A conformal selection set is $\mathcal{S}=\{j: c_j \leq \widehat{Q}(\eta_{nj}^{\rm opt} \mid \bX_{n+j}), 1\leq j \leq m \}$ }
\end{algorithm}
\end{center}

\begin{condition}
\label{condition-Q}
  There exist  a function $ Q(t\mid\bX)$
  and a function class $\mathscr{G} $
such that
(i)
$  P\| \widehat Q(t\mid\bX) - Q(t\mid\bX) \|_1 = o_p(1)$ for each  $t \in (0, 1)$;
(ii)  $\widehat Q(t\mid\cdot), Q(t\mid\cdot)\in \mathscr{G}$ for all $t \in (0, 1)$;
(iii) $ \mathscr{G}$ has finite $\epsilon$-bracket entropy with respect to the $L_1(P)$-norm for any $\epsilon>0$.
\end{condition}

Note that Condition \ref{condition-Q} holds for estimates obtained using the conventional linear quantile regression \citep{koenker2005quantile} and the quantile regression forest \citep{Meinshausen_2006} with mild conditions on $F(y\mid\bx)$ and $\bX$. With this condition, we now establish the asymptotic FDR control for the selection rule given by  Algorithm \ref{algo_cj_no_wt_qrf}.

\begin{theorem}
\label{thm_cj_no_wt_qrf}
Suppose that the calibration data $\mathcal{D}_{\rm ca}$ and the test data $\mathcal{D}_{\rm te}$ are i.i.d., and that Condition \ref{condition-Q} is satisfied. In addition, assume that the varying thresholds $c_j$, $j=1, \dots, m$, are nonrandom constants or random variables that are independent of $\bX_{n+j}$.   For any $ a \in (0, 1)$,  let $ \mathcal{S}$ be the output of Algorithm \ref{algo_cj_no_wt_qrf}.
Then
 $$
\limsup_{n\rightarrow \infty}\e\left[
\frac{ \sum_{j=1}^m I\{j \in \mathcal{S}, Y_{n+j} \leq c_j \}}{ 1 \vee |\mathcal{S}|} \right] \leq a.
$$
\end{theorem}

Theorem \ref{thm_cj_no_wt_qrf} implies that the proposed selection method maintains asymptotic FDR control, though it may slightly underuse the target FDR quota. As discussed previously, the asymptotic FDR approaches the target level quickly as $m$ increases.

\section{Conformal Selection Under Covariate Shift}

\subsection{Weighted Conformal Selection}

In previous sections, we assumed that units in the test samples $\mD_\textrm{te}$ follow the same distribution as the calibration data $\mD_\textrm{ca}$. However, this assumption may not hold in practice. In early-phase drug discovery, for example, experiments are costly and typically limited to a highly selective subset of compounds. As a result, the training dataset consists of previously tested molecules whose clinically relevant properties have been evaluated in experiments.  In contrast, test samples might include novel compounds with characteristics that differ from those in the training data. The shift in the predictor distribution can compromise the validity of the proposed conformal selection algorithm, potentially undermining its utility in identifying promising drug candidates. A similar issue arises in college admissions, where the characteristics of admitted students in the training data may not reflect the broader applicant pool, leading to poor performance of predictive models on new applicants.

Let $f_\textrm{ca}(\bx,y)$ and $f_\textrm{te}(\bx,y)$ denote the joint density functions of $(\bX, Y)$ in $\mD_\textrm{ca}$ and $\mD_\textrm{te}$, respectively. Define the corresponding marginal density functions $g_\textrm{ca}(\bx)=\int f_\textrm{ca}(\bx,y) dy$ and $g_\textrm{te}(\bx)=\int f_\textrm{te}(\bx,y) dy$ and conditional density function  $f_\textrm{ca}(y \mid \bx) = f_\textrm{ca}(\bx, y)/g_\textrm{ca}(\bx)$ and  $f_\textrm{te}(y \mid \bx)= f_\textrm{te}(\bx, y)/g_\textrm{te}(\bx)$. Under covariate shift, the covariate distribution differs between training and testing datasets, that is, $g_\textrm{ca}(\bx)\neq g_\textrm{te}(\bx)$, while the conditional distribution of the outcome given covariates remains the same, that is, $f_\textrm{ca}(y\mid\bx)= f_\textrm{te}(y\mid\bx)=f(y\mid\bx)$. Define $w(\bx)= g_\textrm{te}(\bx)/ g_\textrm{ca}(\bx)$, we have
 \begin{equation*}
\label{covariate-shift1}
\frac{f_\textrm{te}(\bx,y)}{ f_\textrm{ca}(\bx,y)} = \frac{f_\textrm{te}(y \mid \bx)g_\textrm{te}(\bx)}{f_\textrm{ca}(y \mid \bx)g_\textrm{ca}(\bx)}  =w(\bx).
 \end{equation*}
\cite{JinCandes_2023wt} proposed the weighted conformal p-value
 \[
\widetilde p^w_j =\frac{\sum_{i=1}^n w(\bX_i) I(V_i < \widetilde V_{n+j})+
\{w(\bX_{n+j}) +\sum_{i=1}^n w(\bX_i) I(V_i= \widetilde  V_{n+j})\} U_j}{\sum_{i=1}^n w(\bX_i) +w(\bX_{n+j})}.
 \]
This way, the weighted conformal p-value provides a calibrated confidence measure for selecting test units based on their covariates. The authors then applied the Benjamini--Hochberg procedure to $\widetilde p^w_j$'s to control FDR.

The weighted conformal p-value shares the limitations as its unweighted version, notably being more conservative than the oracle conformal p-value. This conservatism is likely to result in low power in accurately selecting the appropriate candidates. Additionally, establishing properties for the weighted conformal p-values presents challenges, as they do not satisfy a key dependence structure known as positive regression dependence on a subset \citep{Barber_2021_AOS}. In what follows, we present a direct FDR control algorithm to optimize the power, while maintaining the FDR under a prespecified threshold.

Under covariate shift, the conditional distribution function of $Y$ given $\bX$, denoted by $F(y\mid \bx)$, is the same across calibration and test data. Under the the null hypothesis $H_j: Y_{n+j}\le c$, the CDF of $\bX_{n+j}$ is given by ${F(c \mid \bx)g_\textrm{te}(\bx)}/ {\int F(c \mid \bu)g_\textrm{te}(\bu) d\bu }$, while under the alternative $H_j^A: Y_{n+j}> c$ it is $ {\bar {F}(c \mid \bx)g_\textrm{te}(\bx)}/{ \int \bar {F}(c \mid \bu)g_\textrm{te}(\bu) d\bu}$. Thus, the likelihood ratio is proportional to $R(\bx; c) =F (c\mid \bx)/\bar  F(c\mid \bx)$, that is, the same likelihood ratio in the absence of covariate shift.
When $m \rightarrow\infty$, the limits of FDP and power in the test sample can be reexpressed using random variables involved in $\mD_\textrm{ca}$; specifically, for $1\le i\le n$,
 \begin{align*}
\lim_{m\rightarrow\infty} \mbox{FDP} (\eta;c)
&=& \frac{ \Pr\{R(\bX_{n+j}; c) \leq \eta, Y_{n+j}\le c\}}{\Pr\{R(\bX_{n+j}; c) \leq \eta\}}
=
\frac{\e\left[ w(\bX_i) I\{ R(\bX_i;c) \leq \eta, Y_i \leq c \}  \right]}{\e\left[ w(\bX_i) I\{R(\bX_i) \leq \eta \} \right]},
\\
\lim_{m\rightarrow\infty} \mbox{Power} (\eta;c)
&=& \frac{ \Pr\{R(\bX_{n+j}; c) \leq \eta, Y_{n+j}> c\}}{\Pr(Y_{n+j}>c)}
=
\frac{ \e\left[ w(\bX_i) I\{R(\bX_i;c) \leq \eta, Y_i > c\}  \right]}{ \e\{ w(\bX_i) I(Y_i > c)  \}}.
 \end{align*}
It is easy to see that these limits can be approximated using $\mD_\textrm{ca}$. For the time being, assume that the covariate shift $w(\bx)$ is known. With a pretrained model $F(y\mid \bx)$, the numerator and denominator in the limit of FDR can be estimated by $n^{-1}\sum_{k=1}^n w(\bX_k) I\{R(\bX_k;c) \leq \eta, Y_k \le c\}$ and $n^{-1}\sum_{k=1}^n  w(\bX_k) I\{R(\bX_k;c) \leq \eta \} $, respectively. Similarly,  the numerator and denominator in the limit of the power function can be estimated by $n^{-1}\sum_{k=1}^n w(\bX_k) I\{R(\bX_k;c) \leq \eta, Y_k > c\}$ and $n^{-1} \sum_{k=1}^n w(\bX_k) I(Y_k > c)$, respectively.

We proposed to optimize the selection rule by  maximizing the weighted power function
 $$
\Psi_n^w(\eta;c) = \frac{\sum_{i=1}^n w(\bX_i) I\{R(\bX_i;c) \leq \eta, Y_i > c\} }{1\vee \sum_{i=1}^n w(\bX_i) I(Y_i > c) }
 $$
subject to the weighted error function
 $$
\zeta_n^w(\eta;c) = \frac{\sum_{i=1}^n w(\bX_i) I\{ R(\bX_i;c) \leq \eta, Y_i \leq c \}}{1\vee \sum_{i=1}^nw(\bX_i) I\{R(\bX_i;c) \leq \eta\} } \leq a.
$$
This essentially replaces the error function  $\zeta_n(\eta;c)$ and the power function $\Psi_n(\eta;c)$ in Algorithm \ref{algo_c_no_wt} with their corresponding weighted version $\zeta_n^w(\eta;c)$ and $\Psi_n^w(\eta;c)$, thereby accommodating the covariate shifts observed between the training and test data.

As discussed in the previous sections, given the calibration data  $\mD_\textrm{ca}$ and the covariate shift $w(\bx)$, both $\zeta_n^w(\eta;c)$ and $\Psi_n^w(\eta;c)$ are step functions of $\eta$ with jumps at $\eta_i=R(\bX_i; c)$, $i=1, \ldots, n$. This property simplifies the process of identifying the optimal critical value for selection. Let $E_n^w =\{\eta_i: \zeta_n^w(\eta_i; c)\le a, 1\leq i \leq n\}$ represent a candidate set of critical values that satisfy the FDR constraint.  We set the optimal critical value $\eta_n^{w,\rm opt}$ to be the one in $E_n^w$ that maximizes the weighted power function so that
$
\eta_n^{w, \rm opt}=\argmax_{\eta\in E_n^w} \Psi_n^w(\eta; c).
$

Furthermore, when the threshold $c_j$ varies across units, the optimal selection rule can be determined by setting $\eta_{nj}^{w,\rm opt}$ as the maximizer of $\Psi_n^w(\eta;c_j)$, subject to the FDR constraint $\zeta_n^w(\eta;c_j) \leq a$. Equivalently, define $\eta_{nj}^{w,\rm opt} =  \argmax_{\eta \in E_{nj}^w} \Psi_n^w(\eta;c_j),$ where $E_{nj}^w = \{\eta_{ij}:  \zeta_n^w(\eta_{ij}; c_j) \leq  a,  1\leq i\leq n \}$ and $\eta_{ij} = R(\bX_i; c_j)$ for $ i=1, \ldots, n$ and $j=1, \ldots, m$. Then the proposed conformal selection set is given by $\mathcal{S} = \{j:  R(\bX_{n+j};c_j) \leq \eta_{nj}^{w,\rm opt}, 1\leq j\leq m \}$. The proposed method is outlined in Algorithm \ref{algo_wt}.

\begin{center}
\begin{algorithm}[h!]
\caption{Weighted Conformal Selection}
\label{algo_wt}

\KwIn{Calibration data $\mD_\textrm{ca} = \{ (\bX_i, Y_i): 1\leq i \leq n \}$, test covariate data $\{\bX_{ j}: 1\leq j \leq m\}$, thresholds $\{c_j\}_{j=1}^m$, target FDR $a \in (0,1)$, a parametric model $F(y\mid\bx;\btheta)$, a pretrained  model $\widehat{w}(\bx)$ for $w(\bx)$, an error function $\zeta_n^w (\eta; \btheta, c_j)$, a power function $\Psi_n^w(\eta; \btheta, c_j)$, and an estimate $\widehat{\btheta}_n$ for $\btheta$ pretrained on a training set}

\KwOut{Selection set $\mathcal{S}$}

\For{$j=1$ \KwTo $m$} {

For $1\leq i \leq n$, obtain $\eta_{ij} = R(\bX_i; \widehat{\btheta}_n, c_j) = F(c_j\mid \bX_{ i};\widehat{\btheta}_n) / \{1-F(c_j\mid \bX_{ i};\widehat{\btheta}_n)\} $.

Obtain a candidate set $E_{nj}^w = \{\eta_{ij}: \zeta_n^w(\eta_{ij};\widehat{\btheta}_n, c_j) \leq  a, 1\leq i \leq n\}$.

If $E_{nj} \neq \varnothing$, find $\eta_{nj}^{w,\rm opt} = \argmax_{\eta \in E_{nj}} \Psi_n^w(\eta;\widehat{\btheta}_n, c_j)$; otherwise set $\eta_{nj}^{w,\rm opt} =  -1 $.
}

\KwResult{A conformal selection set is $\mathcal{S}=\{j:  R(\bX_{n+j}^*;\widehat{\btheta}_n, c_j) \leq \eta_{nj}^{w,\rm opt},  1\leq j\leq m \}$ }
\end{algorithm}
\end{center}
\vspace{-1.5cm}

\subsection{Application to Observational Studies}

As in \cite{JinCandes_2023wt}, we apply the proposed method to select individuals who can potentially benefit from treatment under the Neyman--Rubin potential outcomes framework \citep{Neyman_1923, Rubin_1974}. Let $Y(1)$ and $Y(0)$ represent the potential outcomes under active and control treatment, respectively, with $D=1$ indicating active treatment and $D=0$ otherwise. Under the standard Stable Unit Treatment Value Assumption, we observe $Y=DY(1)+ (1-D)Y(0)$, but not both $Y(0)$ and $Y(1)$ simultaneously. The random vector $\bX$ represents a $p$-dimensional vector of covariates potentially correlated with the treatment received and the potential outcomes.

Assume that $\{(Y_i(0), Y_i(1),  \bX_i, D_i): i=1, \ldots, N\}$ are $N$ i.i.d.\ copies of $(Y(0), Y(1), \bX, D)$, with the first $n=\sum_{i=1}^N D_i$ individuals receiving active treatment and the remaining  $m=N-n$ individuals receiving standard care (control arm). We set the calibration data as $\mathcal{D}_{\rm ca}=\{(\bX_i, Y_i(1)): i=1, \ldots, n\}$ and the test samples as  $\{(\bX_{n+j}, Y_{n+j}(0)): j=1, \ldots, m\}$. Assuming higher outcomes are preferred, our goal is to identify control individuals who might achieve better outcomes if given treatment. This task, however, is challenging as it requires knowledge about the joint distribution of $\{Y_i(0), Y_i(1)\}$, which are correlated but cannot be observed simultaneously. As noted by \cite{JinCandes_2023wt}, their conformal p-value-based approach compares  $Y_{n+j}(0)$ for a control subject to the (weighted) marginal distribution of $Y_i(1)$ from the calibration data, without considering the association between  $Y_{n+j}(0)$ and $Y_{n+j}(1)$. Such strategy can be interpreted as introducing a hypothetical sample of $m$ individuals, where the $j$th individual has an unobserved outcome $Y_{n+j}^*$ and covariate $\bX_{n+j}^*$. We assume $\bX_{n+j}^*=\bX_{n+j}$ and that $(Y_{n+j}^*, \bX_{n+j}^*)$ share the same joint distribution as $(Y_i(1), \bX_i)$ but is independent of $(Y_i(1), Y_i(0), \bX_i)$ in the calibration dataset. The task is then to identify individuals in the hypothetical control arm for whom $Y_{n+j}^*>Y_{n+j}(0)$. Specifically, we test the random null hypothesis $H_j: Y_{n+j}^*\le c_j$ against the alternative $H_j^A: Y_{n+j}^* > c_j$ for $j=1, \ldots, m$, with a varying threshold  $c_j=Y_{n+j}(0)$.

We assume strong ignorability, ${Y(0), Y(1)} \perp D \mid \bX$, which, as shown by \cite{RosenbaumRubin_1983}, is equivalent to ${Y(0), Y(1)} \perp D \mid \pi(\bX)$, where $\pi(\bx) = \Pr(D = 1 \mid \bX = \bx)$ is the propensity score. Let $f(y \mid \bx)$ be the conditional density of $Y(1)$ given $\bX$, and let $g(\bx)$ be the marginal density of $\bX$.  Denote by $h_0(\bx, y)$ and $h_1(\bx, y)$ the conditional densities of $\{\bX, Y(1)\}$ given $D=0$ and $D=1$, respectively. We have
\begin{equation*}
h_1(\bx, y)  = \frac{\pi(\bx) f(y\mid \bx) g(\bx) }{\Pr(D=1)}, \quad
h_0(\bx, y)  =  \frac{\{1-\pi(\bx)\} f(y\mid \bx) g(\bx) }{\Pr(D=0)}.
 \end{equation*}
It is easy to see that the covariate distribution differs between treatment and control groups and that the covariate-shift condition holds as $w(\bx) = \int h_1(\bx, y) dy/\int h_0(\bx, y) dy =  \pi(\bx)  / \{1-\pi(\bx)\} \times \Pr(D=0)/\Pr(D=1) \propto \pi(\bx)  / \{1-\pi(\bx)\}$ depends only on $\bx$.
The propensity score $\pi(\bx)$ can be estimated using pooled covariates from both calibration and test data. A common choice is logistic regression $\rm{logit}\{\pi(\bx)\} = \bbeta^\T \widetilde\bx$, where $\widetilde\bx=(1, \bx^\T)^\T$ and $\bbeta$ is a $(p+1)$-dimensional vector of parameters. The covariate-shift function can be estimated by $\widehat w(\bx) = \exp(\widehat\bbeta^\T\widetilde\bx) \cdot (N-n)/n $, where $\widehat\bbeta$ is the maximum likelihood estimate of $\bbeta$. Alternatively, nonparametric methods or machine learning models like random forests or neural networks can provide a consistent estimate $\widehat w(\bx)$ of $w(\bx)$.

As the threshold $c_j=Y_{n+j}(0)$ varies across individuals $j=1, \ldots, m$, we determine the unit-specific critical value by maximizing the weighted power function
 $$
\Psi_n^w(\eta;\widehat{\btheta}_n, c_j) = \frac{ \sum_{i=1}^n \widehat w(\bX_i) I\{
R(\bX_i; \widehat{\btheta}_n,c_j)
\leq \eta, Y_i(1) >c_j \}}{1\vee \sum_{i=1}^n\widehat  w(\bX_i)  I\{Y_i(1)> c_j\} }
$$
with respect to $\eta$, subject to the constraint:
$$
\zeta_n^w (\eta; \widehat{\btheta}_n, c_j) =
\frac{\sum_{i=1}^n\widehat  w(\bX_i) I\{R(\bX_i; \widehat{\btheta}_n, c_j)\leq \eta,Y_i(1)\leq c_j\}}{1\vee \sum_{i=1}^n\widehat  w(\bX_i) I\{ R(\bX_i; \widehat{\btheta}_n, c_j)\leq \eta\}}\leq  a.
$$
Since $\Psi_n^w(\eta;\widehat{\btheta}_n, c_j)$ is increasing in $\eta$, the optimal critical value is  $\eta_{nj}^{\rm opt} = \max\{\eta_{ij}:  \zeta_n^w (\eta_{ij}; \widehat{\btheta}_n, c_j) \leq a,  1\leq i\leq n \}$, with $\eta_{ij} = R(\bX_{ i}; \widehat{\btheta}_n, c_j)$.  The proposed selection rule can be obtained by modifying Algorithm \ref{algo_wt}, substituting $Y_i$ with $Y_i(1)$, $i=1,\ldots, n$, and $c_j$ with $Y_{n+j}(0)$, $j=1, \ldots, m$.  Theorem \ref{thm-fdr-cs-vc} summarizes its asymptotic properties under Condition \ref{condition-w}.

\begin{condition}
\label{condition-w} Let $P_\textrm{ca}$  denote the probability measure induced by  the distribution of the random variables in $\mathcal{D}_{\rm ca}$. Assume that (i) $ P_\textrm{ca}  \| \widehat w(\cdot) - w(\cdot) \|   = o_p(1)$; (ii) $\widehat w(\cdot)\in  \mathscr{W}$, where $ \mathscr{W}$ is a Glivenko--Cantelli class of functions; and (iii) there exists a function $K (\bx)$ such that $\widetilde w(\bx) \leq K (\bx),  \bx \in \mathcal{X}$ for all $\widetilde w(\cdot) \in  \mathscr{W} $ and $ P_\textrm{ca} K (\bX) <\infty$.
\end{condition}

\begin{theorem}
\label{thm-fdr-cs-vc}
Suppose both the calibration data $\mathcal{D}_{\rm ca} = \{(\bX_i, Y_i(1)): 1\leq i\leq n\}$ and the test sample $\mathcal{D}_{\rm te} = \{(\bX_{n+j}, Y_{n+j}(0)): 1\leq j\leq m \}$ consist of i.i.d.\ random variables. Let $c_j = Y_{n+j}(0),  1\leq j\leq m$.  For a hypothetical sample of $m$ individuals, where the $j$th individual, $j=1, \ldots, m$, has an unobserved outcome $Y_{n+j}^*$ and covariate $\bX_{n+j}^*$. We assume that $\bX_{n+j}^*$ share the same value as $\bX_{n+j}$, and that $(Y_{n+j}^*, \bX_{n+j}^*)$ has the same joint distribution  as $(Y_i(1), \bX_i)$ but is independent of $(Y_i(1), Y_i(0), \bX_i)$  in the calibration dataset. If Conditions \ref{condition-F}
and \ref{condition-w} hold,  then for $ a \in (0, 1)$, the output of the proposed method (i.e., substituting $Y_i$ with $Y_i(1)$ and $c_j$ with $Y_{n+j}(0)$ in Algorithm \ref{algo_wt}) satisfies
$$
\limsup_{n\rightarrow \infty}\e\left[
\frac{ \sum_{j=1}^m I\{j \in \mathcal{S}, Y_{n+j}^* \leq c_j \}}{ 1 \vee |\mathcal{S}|}
\right] \leq a.
$$
\end{theorem}

Theorem \ref{thm-fdr-cs-vc} implies that the proposed conformal selection procedure asymptotically controls FDR with random thresholds $c_j=Y_{n+j}(0)$.
Intuitively, our method selects units with covariate values equal to $\bX_{n+j}$ and an independent future outcome value exceeding $Y_{n+j}(0)$. While this approach, like \cite{JinCandes_2023wt}, ranks and selects units based on potential benefit, it does not offer an individualized causal effect interpretation as its FDR $\zeta_n^w(\eta_{ij};\widehat{\btheta}_n, c_j)$ relies on between-unit rather than within-unit comparisons. Simulation results (see Section 5.2 in the main paper and Section 10.5 in the Supplementary Materials) suggest that the proposed methods effectively control the FDR across varying levels of association between the two potential outcomes.

\section{Numerical Studies}

In this section, we compare the performance of the proposed conformal selection methods with competing approaches using both simulated and semi-simulated data. The scenarios encompass both symmetric and skewed outcome distributions, as well as settings with and without covariate shifts between calibration and test samples. Detailed descriptions and results can be found in the Supplementary Materials.

\subsection{Numerical Evaluation Using Simulated Data}

For each simulation, the training data $\mathcal{D}_\textrm{tr} = \{(\bX_i,Y_i):i=1,\ldots, n\}$ and calibration data $\mathcal{D}_\textrm{ca} = \{(\bX_i,Y_i):i=1,\ldots,n\}$ each contain $n = 1000$ samples. Performance is evaluated on test datasets of sizes $m \in\{100,500,1000\}$. The objective is to test $m$ random hypotheses of the form $H_j: Y_{n+j} \leq c_j$, $j=1,\ldots,m$.  Each scenario is repeated 2000 times.

\begin{figure}[ht]
\centering
\caption{Empirical FDP and empirical power for an exchangeable outcome generated under a location mixture model at the 10\% target FDR}
\label{fig_mixture}
\includegraphics[width=0.8\linewidth]{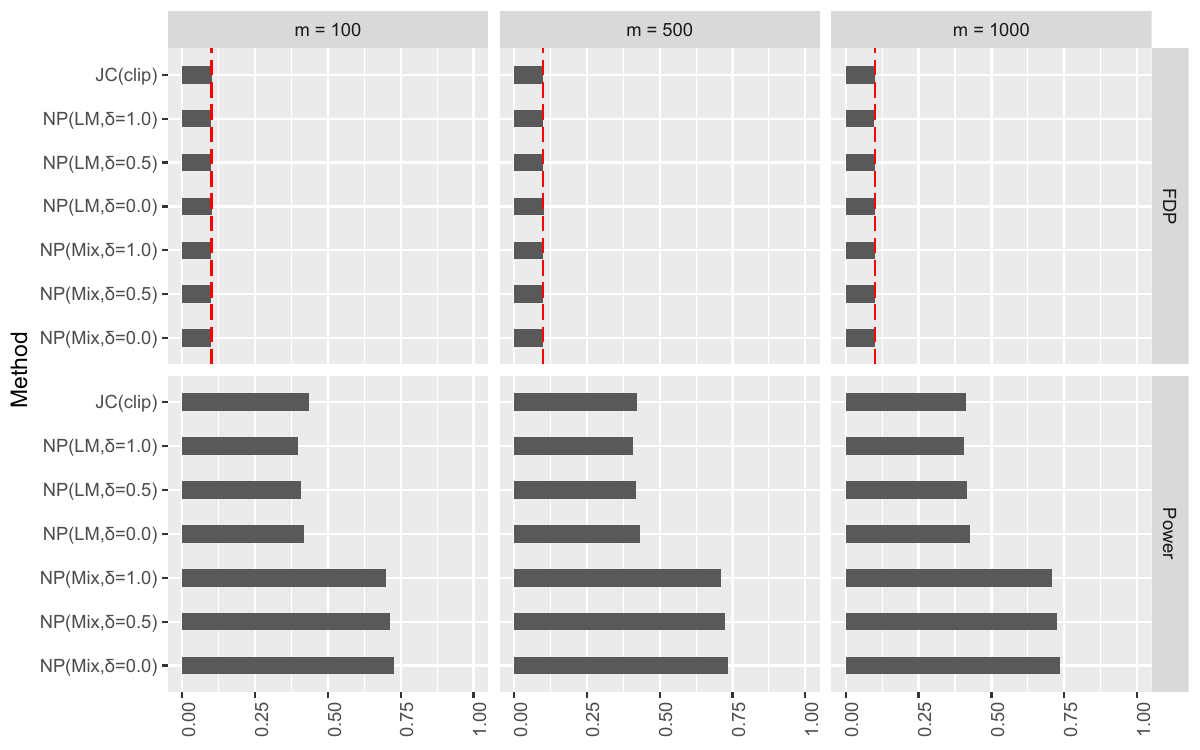}
\end{figure}

In Scenario 1, we evaluate the performance of the proposed likelihood-ratio-based approaches with a mixture model and compare them with competing methods. Specifically, the covariate $X$ is drawn from a standard normal distribution, and the conditional distribution of $Y$ given $X=x$ follows a two-component normal mixture with component means $\mu_1(x)=-1 + x + x^2$ and $\mu_2(x)=1 - 2x$, a common variance of 2.25, and mixing probabilities of 0.4 and 0.6.
Three different methods are compared: the first two methods implement Algorithm \ref{algo_c_no_wt} with different working models, while the third applies the cfBH procedure with a clipped score as proposed in \cite{JinCandes_2023}.
Specifically, the first method, NP(Mix), inspired by the Neyman--Pearson paradigm, fits a two-component mixture regression model that includes the true model as a special case \citep{mixtools}, while the second and the third method, denoted respectively by NP(LM) and JC(clip), fit a linear regression model with covariates $X$ and $X^2$ and thus a misspecified working model. We set a constant threshold of $c_j\equiv -2$ and apply the modified error function \eqref{zeta-n-plus} with $\delta=0$, $0.5$ and $1.0$. Figure \ref{fig_mixture} summarizes empirical FDP and power at a 10\% target FDR for each method and sample size. The results show that JC(clip) effectively controls FDR control but yields low power (around 40\%--43\%). NP(LM) shows comparable performance to JC(clip), as both fit the same (misspecified) working model. In contrast, NP(Mix), which fits a working model that contains the truth, achieves significantly higher power (around 70\%--74\%) while yielding a slightly elevated FDP. Moreover, applying the $\delta$-adjustment to the error function slightly reduces FDP, providing a better balance between error control and power.

Scenario 2 evaluates the proposed methods with approximately symmetric outcome distributions. Following \cite{JinCandes_2023}, each component of $\bX = (X_1, \ldots, X_{20})$ is independently drawn from $U[0, 1]$, and we generate $Y = \mu(\bX) + \epsilon(\bX)$, where  $\mu(\bX)=5(X_1 X_2 + e^{X_4 - 1})$ and $\epsilon(\bX)$ is independently normally distributed with mean 0 and one of three variance functions:  $2.25$, $(5.5-|\mu(\bx)|)/2$, and $0.25 \mu(\bx)^2 I\{|\mu(\bx)|<2\} + 0.5|\mu(\bx)|I\{|\mu(\bx)|\geq 1\}$. Both constant thresholds  ($c_j \equiv 0$) and varying thresholds ($c_j \sim U[0,1]$) are considered. Three proposed methods are implemented: Algorithm \ref{algo_c_no_wt} (constant thresholds), Algorithm \ref{algo_cj_no_wt} (varying thresholds), and Algorithm \ref{algo_cj_no_wt_qrf}, which employs quantile regression and applies to both scenarios. The first two methods, collectively referred to as NP(RF), estimate ${\mu}(\bX)$ via random forests, assuming normally distributed residuals. The third method, denoted NP(qRF), employs quantile regression forests (qRF) to estimate the outcome quantiles across the candidate set  $\{0.1, 0.11, \ldots, 0.9\}$. These methods are compared with the cfBH procedure using residual and clipped scores, referred to as JC(res) and JC(clip), where  random forests are used to estimate the mean functions. Simulation results (see Figure S1 in the Supplementary Materials) show that NP(RF) and JC(clip) achieve comparable power across all settings. While JC(clip) offers slightly better exact FDR control compared to NP(RF), the $\delta$-adjustment in \eqref{zeta-n-plus} allows NP(RF) to achieve comparable FDP without sacrificing power. NP(qRF) consistently outperforms in both FDR control and power, particularly in scenarios with greater heteroscedasticity in the outcome distribution.

\begin{figure}[ht]
\centering
\caption{Empirical FDP and empirical power for a covariate-shifted outcome generated under Scenario 2 with variance function $\sigma_2^2(\bx)$ at the 10\% target FDR}
\label{fig_scenario3}
\includegraphics[width=0.8\linewidth]{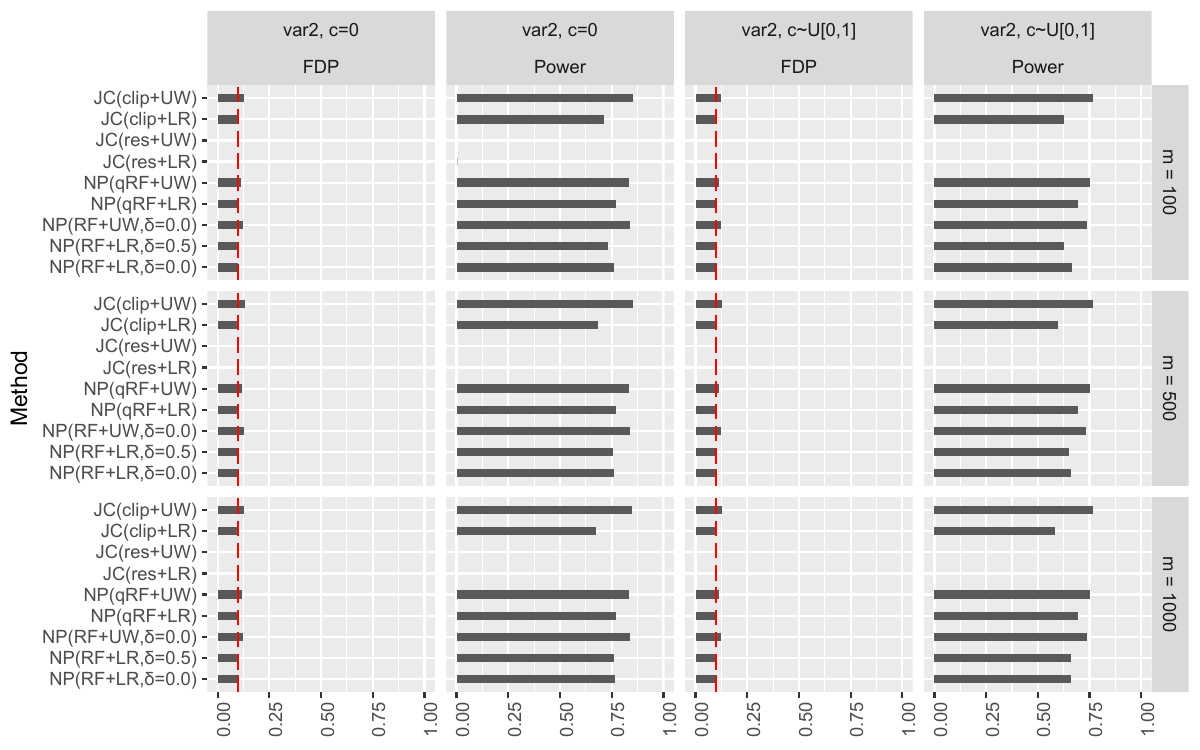}
\end{figure}

In Scenario 3, we assess the performance of proposed methods under covariate shift. Data are generated under Scenario 2 with variance function $(5.5-|\mu(\bx)|)/2$ and tested with the same thresholds therein. Here we use a logistic regression model to create covariate shift between test set and training/calibration set. Specifically, the generated $n$ subjects of training/calibration data and $m$ subjects of test data satisfy $f_\textrm{te}(\bx,y) \propto w(\bx)f_\textrm{ca}(\bx,y)$, where $w(\bx) = \exp(-x_1 + x_2 -x_3)$. We estimate the covariate-shift function using a correctly specified logistic regression (LR). We evaluate the performance of our unweighted (UW) methods and their weighted counterparts, and compare them to the weighted conformal selection procedure with heterogeneous pruning \citep[Algorithm 1 in]{JinCandes_2023wt}. As expected, unweighted methods fail to control FDR under covariate shift (Figure \ref{fig_scenario3}). Among the weighted methods, all perform well except for JC(res), with our proposed methods slightly outperforming the clipped score method.

\begin{figure}[hp]
\centering
\caption{Empirical FDP and empirical power for an outcome generated under Scenario 4 at the 10\% target FDR}
\label{fig_scenario4}

\begin{subfigure}{0.7\textwidth}
\centering
\includegraphics[width=\linewidth]{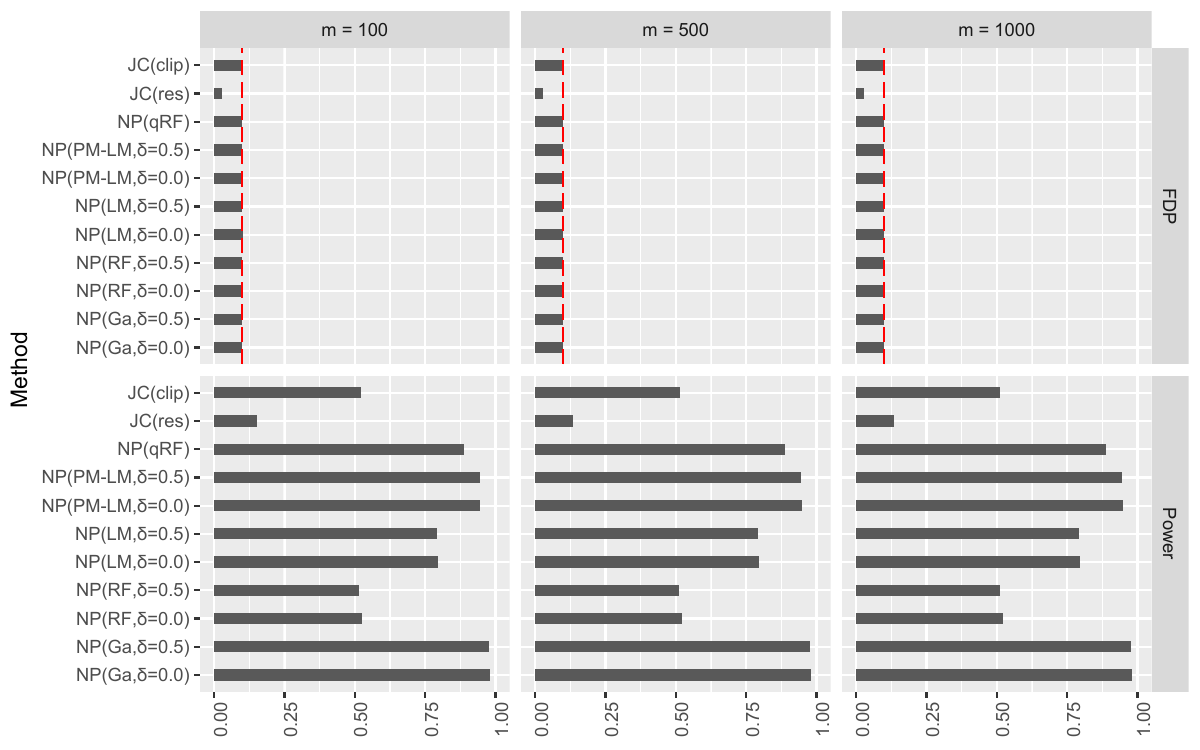}
\caption{Exchangeable outcome with low-dimensional covariate}
\end{subfigure}

\begin{subfigure}{0.7\textwidth}
\centering
\includegraphics[width=\linewidth]{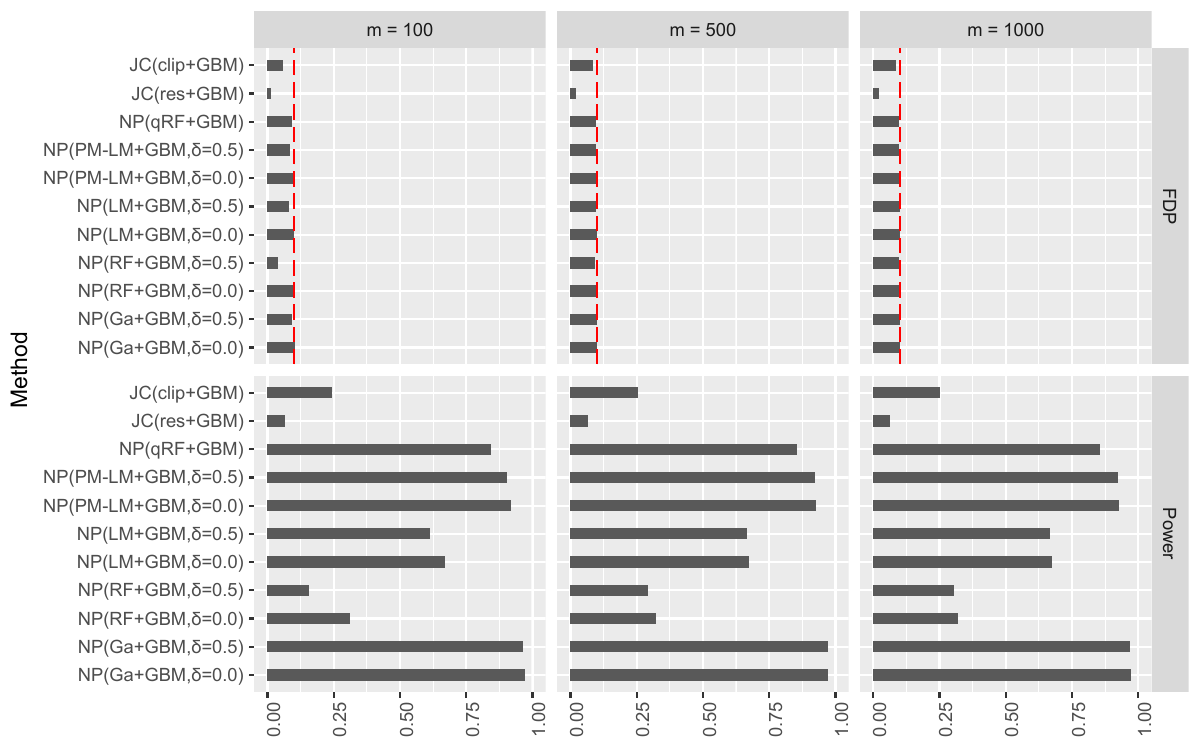}
\caption{Covariate-shifted outcome with low-dimensional covariate}
\end{subfigure}

\begin{subfigure}{0.7\textwidth}
\centering
\includegraphics[width=\linewidth]{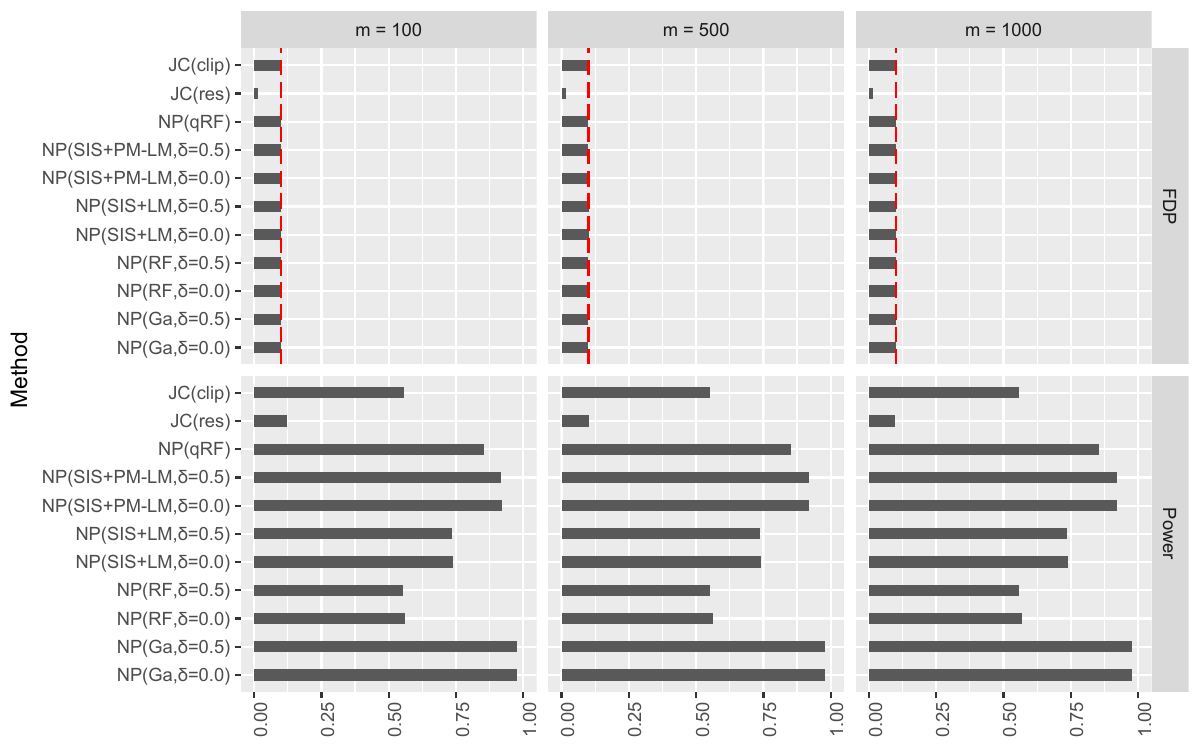}
\caption{Exchangeable outcome with high-dimensional covariate}
\end{subfigure}

\end{figure}

Scenario 4 examines right-skewed outcomes with and without covariate shifts. We generate $\bX = (X_1, \ldots, X_{5})$, with each component independently drawn from $U[0, 2]$, and $Y$ from a Gamma distribution with shape parameter 5 and scale parameter $\exp(\bbeta^\T \widetilde{\bX} -2 X_1^2) / 5$, where $\widetilde{\bX} = (1, \bX^\T)^\T$ and $\bbeta = (1, -1, 1, 2, -1, 1)^\T$. Five working models are considered: (1) a correctly specified Gamma model (Ga) estimated using the standard maximum likelihood method; (2) a location-shift model  $Y = \mu(\bX) + \epsilon$, where $\mu(\cdot)$ is trained by RF and  $\epsilon \sim N(0, \sigma_\epsilon^2)$ estimated from residuals; (3) a linear model (LM) with predictors effects for $\bX$ estimated via the standard maximum likelihood method; (4) the power maximization (PM) procedure described in Section \ref{max-power-model-mis} under the working LM; and (5) a quantile regression model trained by qRF. Here we consider constant thresholds $c_j\equiv 5$. To induce covariate shift, the test sample is selected with probability proportional to $\{1+B_{24}(x_1)\}/4$, where $B_{24}(\cdot)$ is the CDF of the $\textrm{Beta}(2,4)$ distribution.
The covariate-shift function is estimated using the gradient boosting machine \citep[GBM]{Friedman_2001}.
Simulation results summarized in Figure \ref{fig_scenario4}(a) and Figure \ref{fig_scenario4}(b) show that all methods adequately control FDR. Employing the true Gamma working model consistently achieves the highest power across all settings. NP(RF) and JC(clip) show comparable FDPs but much lower power, while the proposed power maximization method NP(PM-LM) improves power by 20\% over NP(LM) with a ``plug-in" working LM. Under covariate shift, Gamma models with GBM weighting achieve high power and maintain FDP control. Weighted methods, including NP(PM-LM+GBM), generally outperform JC methods in power, with NP(PM-LM+GBM) yielding a 37\% power improvement over NP(LM+GBM).

Finally, we explore a high-dimensional setting where data are generated from the same Gamma distribution, but with $\bX\sim U[0,2]^{20}$ and $\bbeta = (1, -1, 1, 2, -1, 1, \bm{0}_{15})^\T$. The SIS procedure \citep{FanLv_2008_SIS,SaldanaFeng_2018} with a Gaussian family is first applied to identify covariates strongly correlated with the outcome. The selected covariates are then used to train a linear model (SIS+LM). As shown in Figure \ref{fig_scenario4}(c), the high-dimensional results closely align with the low-dimensional case. Notably, the proposed power maximization method achieves a 25\% improvement in power compared to NP(LM). Even though misspecified models are used in both the screening process and the outcome regression, the proposed method maintains robust performance. Across all scenarios, JC methods maintain strict FDR control but generally have lower power.

\subsection{Performance Evaluation Using Semi-Simulated Data}

We apply the proposed methods to semi-simulated data derived from the National Study of Learning Mindsets (NSLM), a randomized trial of a growth mindset intervention in school children. To emulate an observational study, \cite{carvalho2019assessing} generated a synthetic dataset from NSLM by adding confounding in treatment selection while maintaining original data structure and effect sizes. The dataset was further split into  $\mZ_1$ (2079 subjects) and $\mZ_2$ (8312 subjects). Following the potential outcome and propensity score generation processes outlined in Section 4.4 of \cite{lei2021conformal}, we use $\mZ_1$ to estimate $\widehat{m}_d(\bx)$ and $\widehat{r}_d(\bx)$ for $d = 0,1$ and generate potential outcomes for subjects in $\mZ_2$ as follows: $Y(1) = \widehat{m}_0(\bX) + \tau(\bX) + \tau_0 + 0.5\widehat{r}_1(\bX)\epsilon_{1}$ and $Y(0) = \widehat{m}_0(\bX) + 0.5\widehat{r}_0(\bX)\epsilon_{0}$. Here, $\tau(\bx)$ is defined as in Equation (1) of \cite{carvalho2019assessing} and $(\epsilon_{0},\epsilon_{1})$ follow a bivariate normal distribution with mean zero, variance 2, and covariance $r$.  The treatment propensity model, estimated using random forest on  $\mZ_1$ and applied to $\mZ_2$, results in about 29\% treated subjects. For each of 2,000 simulations, three datasets of 1,000 subjects are randomly sampled from  $\mZ_2$: the training set, $\mD_\textrm{tr}$, keeps all 1000 sampled subjects; the calibration set, $\mD_\textrm{ca}$, includes only  treated subjects; and the testing set, $\mD_\textrm{te}$, retains only control subjects. The threshold $c_j$ for each test sample is set to its control outcome $Y_{n+j}(0)$, $j=1, \ldots, m$. We vary $r$ to examine different correlation levels between $Y(0)$ and $Y(1)$ and vary $\tau_0$ to assess various degrees of overlap between the two potential outcome distributions.

\begin{figure}[hp]
\centering
\caption{Empirical FDP and empirical power for the semi-simulated dataset based on the NSLM study ($\rho$ denotes the correlation between two potential outcomes)}
\label{fig_NSLM}

\begin{subfigure}{0.9\textwidth}
\centering
\includegraphics[width=\linewidth]{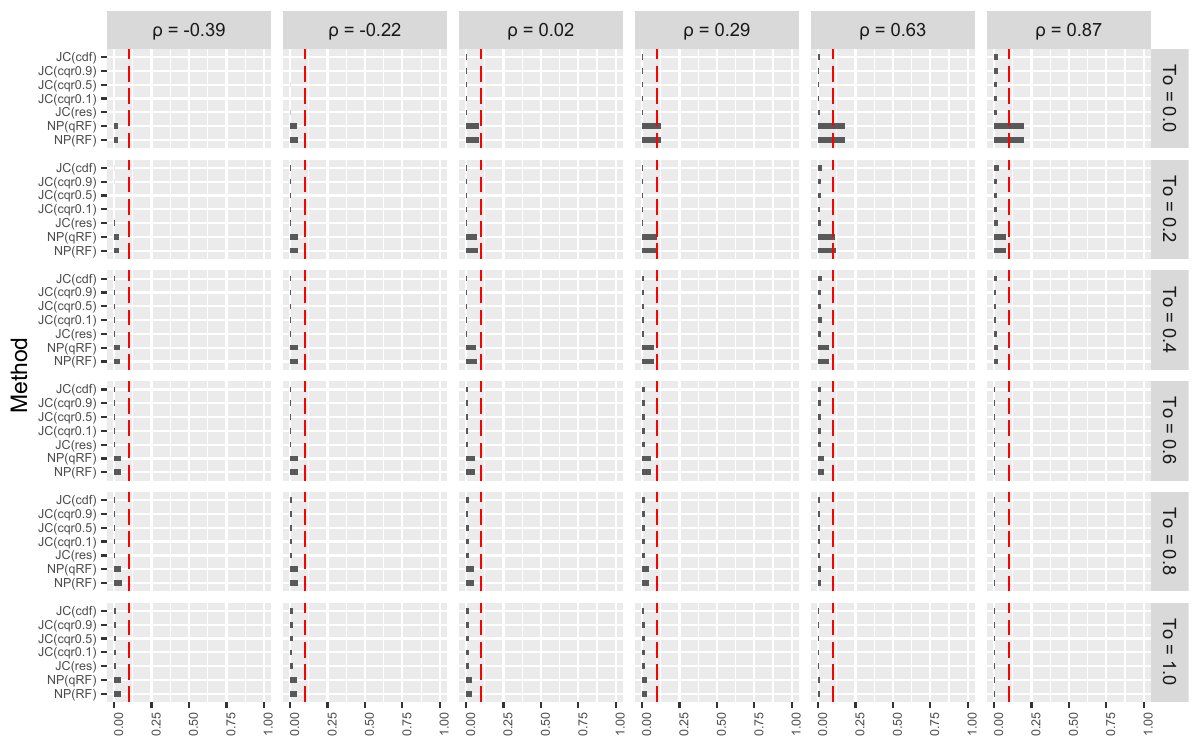}
\caption{Empirical FDP}
\end{subfigure}

\begin{subfigure}{0.9\textwidth}
\centering
\includegraphics[width=\linewidth]{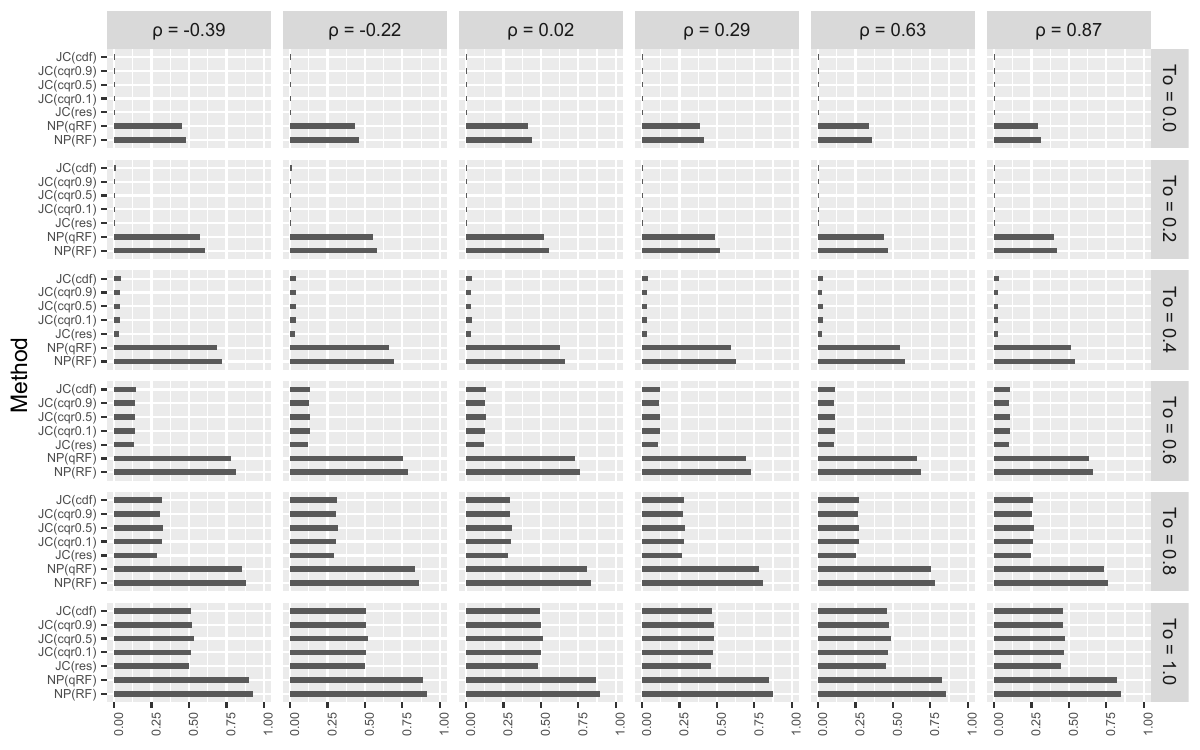}
\caption{Empirical power}
\end{subfigure}

\end{figure}

The outcome distribution is estimated using two approaches: a location-shift working model with the mean function trained via RF, and qRF for full conditional quantile estimation, both based on treated subjects in the training dataset. The propensity score is trained with GBM on all training samples to meet the independence requirement for FDR control of the weighted JC method, though our method does not have this requirement. The proposed weighted methods, NP(RF+GBM) and NP(qRF+GBM), are then compared to the weighted JC method using residual, clipped function, CDF, and conformalized quantile regression (cqr) at the 10th, 50th, and 90th quantiles to define the nonconformity score \citep[see Sections 5.3 and 5.4 in][]{JinCandes_2023wt}.
The simulation results summarized in Figure \ref{fig_NSLM} show that treating $c_j=Y_{n+j}(0)$ as a random threshold, thereby ignoring the correlation between $Y_{n+j}(0)$ and $Y_{n+j}(1)$ in the proposed selection procedures, does not compromise FDR control.
Moreover, the proposed methods outperform the competing methods in terms of power.
Note that, when $\tau_0=0$, our proposed methods have inflated FDP due to near-complete overlap in potential outcome distributions (see Figure S5 in the Supplementary Materials), while JC methods have nearly no power.

\section{Concluding Remarks}

This paper presents new conformal selection methods that maximize selection power directly while controlling the FDR. Existing approaches, such as those proposed by \cite{JinCandes_2023} and \cite{JinCandes_2023wt}, derive conformalized p-values by replacing unobserved future outcomes in the nonconformity score with pre-specified thresholds, which violates exchangeability between test and calibration samples. Our approach reframes the selection problem as a hypothesis test on the distribution of observed covariates. By applying the Neyman--Pearson criterion, we derive likelihood-ratio-based selection rules that achieve asymptotically optimal selection under FDR control and remain robust to model misspecification. Notably, the proposed selection method is versatile and can accommodate alternative definitions of the power function, such as
\begin{equation*}
\Psi_{n}(\eta;c)  = \frac{\sum_{k=1}^n I\{R(\bX_k;c) \ge \eta, Y_k \leq c\}}{1 \vee  \sum_{k=1}^n I\{R(\bX_k;c) \geq \eta\}},
\end{equation*}
allowing it to be tailored to different application needs.

Finally, the proposed methods achieve asymptotically optimal power when the working model is correctly specified. To better approximate the data structure, multiple trained  models, $\widehat\mu_k(\bX)$ under working models $Y =  \mu_k(\bX) + \epsilon_k$ for $k = 1, \dots, K$, can be combined into an ensemble model $\widehat{\mu}(\bX) = \sum_{k=1}^K \omega_k \widehat{\mu}_k(\bX)$, with weights $\omega_k$'s satisfying $\sum_{k=1}^K \omega_k = 1$.  These weights can be estimated by implementing the method in Section \ref{max-power-model-mis} to improve power. Simulation results in Table S5 of the Supplementary Materials demonstrate the feasibility of this ensemble approach, suggesting it merits further study.

\section*{Supplementary material}
The online supplement (available upon request) contains all the technical proofs, details of the simulation results, and R codes to implement the proposed methods.

\section*{Acknowledgements}

Liu's research is supported by
the National Natural Science Foundation of China (12171157, 32030063), Fundamental Research Funds for the Central Universities and the 111 Project (B14019).

\bibliographystyle{natbib}
\bibliography{ConfSelect}

\end{document}